\documentclass[12pt,preprint]{aastex}

\shorttitle{Development of the spectrum} \shortauthors{Qin}

\begin{document}

\title{Development of the spectrum of gamma-ray burst pulses
influenced by the intrinsic spectral evolution and the curvature
effect}
\author{Y.-P. Qin\altaffilmark{1,2}}

\altaffiltext{1}{Center for Astrophysics, Guangzhou University,
Guangzhou 510006, P. R. China; ypqin@gzhu.edu.cn}

\altaffiltext{2}{Physics Department, Guangxi University, Nanning
530004, P. R. China}

\begin{abstract}
Spectral evolution of gamma-ray burst pulses assumed to arise from
emission of fireballs is explored. It is found that, due to the
curvature effect, the integrated flux and peak energy are well
related by a power law in the decaying phase of pulses, where the
index is about 3, being free of the intrinsic emission and the
Lorentz factor. The spectrum of a pulse in its decaying phase
differs slightly for different intrinsic spectral evolution
patterns, indicating that it is dominated by the curvature effect.
In the rising phase, the integrated flux keeps increasing whilst the
peak energy remains unchanged when the intrinsic emission bears an
unchanged spectrum. Within this phase, the flux decreases with the
increasing of the peak energy for a hard-to-soft intrinsic spectrum,
and for a soft-to-hard-to-soft intrinsic spectrum, the flux
generally increases with the increasing of the peak energy. An
intrinsic soft-to-hard-to-soft spectral evolution within a co-moving
pulse would give rise to a pulse-like evolutionary curve for the
peak energy.
\end{abstract}

\keywords{gamma-ray bursts, gamma-rays, relativity}

\section{Introduction}

Recent Swift observations of early X-ray afterglows of gamma-ray
bursts (GRBs) revealed that the previous interpretation of the GRB
afterglow, which is based on the mechanism of external shocks, is
unlikely to be the mechanism accounting for these new findings.
Instead, the internal shock origin and the so-called ``curvature
effect'' might be responsible for many of the early X-ray afterglows
[1-8]. The curvature effect is a combined effect that includes the
delay of time and the shifting of the intrinsic spectrum as well as
other relevant factors of an expanding fireball (see [9] for a
detailed explanation). The effect was widely investigated in recent
years in both the prompt emission and early X-ray afterglow of GRBs
[4], [7-9], [10-23].

Early study of the spectral evolution of GRBs revealed that the
general evolution of the spectra of these objects is hard-to-soft
(e.g., [24]). This issue was also frequently investigated in recent
years [11], [25-28]. Reference [14] showed that the
hardness-intensity correlation could be accounted for by the
curvature effect. Authors of [9] explored the evolutionary curve of
the hardness ratio (HRC) of GRB pulses and concluded that the main
characteristics of the HRC of some GRB FRED pulses are in agreement
with what predicted by the curvature effect. In addition, they
showed that the curvature effect alone could not completely explain
the observed HRCs. Instead, a hard-to-soft intrinsic spectral
evolution might exist and probably play a role in producing the
observed HRCs.

Here, we investigate how the intrinsic spectral evolution and the
curvature effect combine in producing an observed spectrum arising
from a fireball pulse. The paper is organized as follows. In Section
2, we present formulas that can directly applied to the case of the
intrinsic spectral evolution. Spectral evolution within a pulse of
fireballs radiated with a typical intrinsic emission form is
investigated in various aspects in Section 3. In Section 4, other
intrinsic emission forms are considered. In Section 5, we pay our
attention to various situations of the soft-to-hard-to-soft
intrinsic emission. Influence of the Lorentz factor on the spectral
evolution is investigated in Section 6. Presented in Section 7 is a
signature of the curvature effect shown by the evolutionary curve of
the peak energy in the decaying phase of pulses. Conclusions are
given in Section 8.

\section{General formulas of flux and count rates for expanding fireballs}

Here we present more general formulas compared with those derived in
[13] and [19] so that they could be directly applied to various
cases of the intrinsic spectral evolution pattern. For more general
applications, formulas with various variables are provided.

\subsection{Formulas in terms of the integral of $\protect\theta $}

For a radiation associated with an intrinsic emission independent of
direction, the expected flux of a fireball expanding with a constant
Lorentz factor is [13]
\begin{equation}
f_\nu (t)=\frac{2\pi }{D^2}\int_{\widetilde{\theta }_{\min }}^{\widetilde{%
\theta }_{\max }}I_\nu (t^{\prime },\nu,\nu _0 )R^2(t^{\prime })\cos
\theta \sin \theta d\theta ,
\end{equation}
where $I_\nu (t^{\prime },\nu,\nu_0 )$ is the observer frame
intensity; $\nu $ is the observation frequency; $\nu _0$ is the rest
frame frequency which emits from differential surface $ds_\theta $
and gives rise to $\nu $ (see equation 9 presented below for the
relation between $I_\nu$ and $\nu _0$); $t$ is the observation time;
$D$ is the distance of the fireball to the observer; $\theta $ is
the angle, of $ds_\theta $, of the fireball, to the line of sight;
$t^{\prime }$ is the emission time (in the observer frame), called
local time, of photons which emit from $ds_\theta $; $R(t^{\prime
})$ is the radius of the fireball measured at time $t^{\prime }$.
The integral range of $\theta $, $\widetilde{\theta }_{\min }$ and
$\widetilde{\theta }_{\max }$, will be determined by the concerned
area of the fireball surface as well as the emission ranges of the
frequency and the local time. As shown in [13] and [19], $ t^{\prime
}$ and $t$ are related by
\begin{equation}
t^{\prime }=\frac{t-t_c-\frac Dc+\frac{R_c}c\cos \theta }{1-\beta
\cos \theta }+t_c,
\end{equation}
where $t_c$ and $R_c$ are constants. Assume that the area of the
fireball surface concerned is confined within
\begin{equation}
\theta _{\min }\leq \theta \leq \theta _{\max }
\end{equation}
and the emission time $t^{\prime }$ is confined within
\begin{equation}
t_c\leq t_{\min }^{\prime }\leq t^{\prime }\leq t_{\max }^{\prime },
\end{equation}
and besides them there are no other constraints to the integral
limit of (1). According to (3) and (4), one can verify that the
lower and upper integral limits of (1) could be determined by
\begin{equation}
\widetilde{\theta }_{\min }=\cos ^{-1}\min \{\cos \theta _{\min },\frac{%
t_{\max }^{\prime }-t+\frac Dc}{(t_{\max }^{\prime }-t_c)\beta +\frac{R_c}c}%
\}
\end{equation}
and
\begin{equation}
\widetilde{\theta }_{\max }=\cos ^{-1}\max \{\cos \theta _{\max },\frac{%
t_{\min }^{\prime }-t+\frac Dc}{(t_{\min }^{\prime }-t_c)\beta +\frac{R_c}c}%
\},
\end{equation}
respectively (for a detailed derivation, one could refer to [13] and
[19]. Applying the relation between the radius of the fireball and
the observation time we come to the following form of flux (see [13]
and [19]):
\begin{equation}
f_\nu (t)=\frac{2\pi [(t-t_c-\frac Dc)\beta c+R_c]^2}{D^2}\int_{\widetilde{%
\theta }_{\min }}^{\widetilde{\theta }_{\max }}\frac{I_\nu
(t^{\prime },\nu,\nu _0 )\cos \theta \sin \theta }{(1-\beta \cos
\theta )^2}d\theta .
\end{equation}

From (2) one finds that, for any given values of $t$ and $\theta $,
local time $t^{\prime }$ would be uniquely determined. If this value
of $t^{\prime
}$ is within the range of (4), then there will be photons emitted at $%
t^{\prime }$ from the small surface area of $\theta $ reaching the
observer at $t$ [when $\theta $ is within the range of (3), this
small area would be included in the above integral, otherwise it
would not]. Obviously, for a
certain value of $\theta $, the range of $t$ depends on the range of $%
t^{\prime }$. Inserting (2) into (4) and applying (3) we obtain
\begin{equation}
\begin{array}{l}
(1-\beta \cos \theta _{\min })t_{\min }^{\prime }+(t_c\beta -\frac{R_c}%
c)\cos \theta _{\min }+\frac Dc\leq t \\
\leq (1-\beta \cos \theta _{\max })t_{\max }^{\prime }+(t_c\beta -\frac{R_c}%
c)\cos \theta _{\max }+\frac Dc%
\end{array}
.
\end{equation}
It suggests clearly that observation time $t$ is limited when emission time $%
t^{\prime }$ is limited.

Let $t_0$ be the proper time corresponding to the coordinate time
$t^{\prime }$ and $\nu _0$ be the co-moving frequency corresponding
to $\nu $. The latter two are related by the Doppler effect. It is
well known that the observer frame intensity $I_\nu (t^{\prime
},\nu,\nu_0 )$ is related to the rest frame intensity $I_{0,\nu
}(t_0,\nu _0)$ by
\begin{equation}
I_\nu (t^{\prime },\nu,\nu_0 )=(\frac \nu {\nu _0})^3I_{0,\nu
}(t_0,\nu _0),
\end{equation}
which could be written by
\begin{equation}
I_\nu (t^{\prime },\nu,\nu_0 )=\frac{I_{0,\nu }(t_0,\nu
_0)}{(1-\beta \cos \theta )^3\Gamma ^3}
\end{equation}
when applying the Doppler effect. The flux is now able to be written
by
\begin{equation}
f_\nu (t)=\frac{2\pi [(t-t_c-\frac Dc)\beta c+R_c]^2}{D^2\Gamma ^3}\int_{%
\widetilde{\theta }_{\min }}^{\widetilde{\theta }_{\max
}}\frac{I_{0,\nu }(t_0,\nu _0)\cos \theta \sin \theta }{(1-\beta
\cos \theta )^5}d\theta ,
\end{equation}
where $\widetilde{\theta }_{\min }$ and $\widetilde{\theta }_{\max
}$ are determined by (5) and (6), respectively, $\nu _0$ and $\nu $
are related by the Doppler effect, and $t$ is confined by (8).

As $t_0$ and $t^{\prime }$ represent the same moment, according to
the theory of special relativity, they are related by $ t^{\prime
}-t_c=\Gamma (t_0-t_{0,c})$, where $t_{0,c}$ is a constant (here we assign $t^{\prime }=t_c$ when $%
t_0=t_{0,c}$). Assign $ t^{\prime }=t_{\min }^{\prime
}|_{t_0=t_{0,\min }}$ and $ t^{\prime }=t_{\max }^{\prime
}|_{t_0=t_{0,\max }}$. We get $ t_{\min }^{\prime }=\Gamma
(t_{0,\min }-t_{0,c})+t_c$, and $ t_{\max }^{\prime }=\Gamma
(t_{0,\max }-t_{0,c})+t_c$. Then the constraint of (4) is identical
to
\begin{equation}
t_{0,c}\leq t_{0,\min }\leq t_0\leq t_{0,\max }.
\end{equation}

We then get from (5)-(6) that
\begin{equation}
\widetilde{\theta }_{\min }=\cos ^{-1}\min \{\cos \theta _{\min },\frac{%
(t_{0,\max }-t_{0,c})\Gamma +\frac Dc-(t-t_c)}{(t_{0,\max
}-t_{0,c})\Gamma \beta +\frac{R_c}c}\}
\end{equation}
and
\begin{equation}
\widetilde{\theta }_{\max }=\cos ^{-1}\max \{\cos \theta _{\max },\frac{%
(t_{0,\min }-t_{0,c})\Gamma +\frac Dc-(t-t_c)}{(t_{0,\min
}-t_{0,c})\Gamma \beta +\frac{R_c}c}\}.
\end{equation}
In addition, from (2) and $ t^{\prime }-t_c=\Gamma (t_0-t_{0,c})$ we
obtain
\begin{equation}
t_0=\frac{t-t_c-\frac Dc+\frac{R_c}c\cos \theta }{(1-\beta \cos
\theta )\Gamma }+t_{0,c}
\end{equation}
and from $ t_{\min }^{\prime }=\Gamma (t_{0,\min }-t_{0,c})+t_c$, $
t_{\max }^{\prime }=\Gamma (t_{0,\max }-t_{0,c})+t_c$, and (8) we
gain
\begin{equation}
\begin{array}{l}
(1-\beta \cos \theta _{\min })[(t_{0,\min }-t_{0,c})\Gamma +t_c]+(t_c\beta -%
\frac{R_c}c)\cos \theta _{\min }+\frac Dc\leq t \\
\leq (1-\beta \cos \theta _{\max })[(t_{0,\max }-t_{0,c})\Gamma
+t_c]+(t_c\beta -\frac{R_c}c)\cos \theta _{\max }+\frac Dc%
\end{array}
.
\end{equation}

When calculating the flux from (11), we need to convert variable $t_0$ to $%
\theta $ via (15) and convert $\nu _0$ to $\theta $ according to the
Doppler effect $ \nu _0=(1-\beta \cos \theta )\Gamma \nu $.

Light curves of gamma-ray bursts are always presented in terms of
count rates within an energy range. The count rate within energy
channel $[\nu _1,\nu _2]$ is determined by
\begin{equation}
\frac{dn(t)}{dt}=\int_{\nu _1}^{\nu _2}\frac{f_\nu (t)}{h\nu }d\nu .
\end{equation}
Inserting (11) leads to
\begin{equation}
\frac{dn(t)}{dt}=\frac{2\pi [(t-t_c-\frac Dc)\beta c+R_c]^2}{D^2\Gamma ^3h%
}\int_{\widetilde{\theta }_{\min }}^{\widetilde{\theta }_{\max
}}[\int_{\nu _1}^{\nu _2}\frac{I_{0,\nu }(t_0,\nu _0)}\nu d\nu
]\frac{\cos \theta \sin \theta }{(1-\beta \cos \theta )^5}d\theta .
\end{equation}
In the case that $\theta $ could be treated as a constant, from  $
\nu _0=(1-\beta \cos \theta )\Gamma \nu $ one gets $ \frac{d\nu }\nu
|_{\theta =const}=\frac{d\nu _0}{\nu _0}$. Replacing variable $\nu $
with $\nu _0$ we get from (18) that
\begin{equation}
\frac{dn(t)}{dt}=\frac{2\pi [(t-t_c-\frac Dc)\beta c+R_c]^2}{D^2\Gamma ^3h%
}\int_{\widetilde{\theta }_{\min }}^{\widetilde{\theta }_{\max
}}[\int_{\nu _{0,1}}^{\nu _{0,2}}\frac{I_{0,\nu }(t_0,\nu _0)}{\nu
_0}d\nu _0]\frac{\cos \theta \sin \theta }{(1-\beta \cos \theta
)^5}d\theta
\end{equation}
where $ \nu _{0,1}=(1-\beta \cos \theta )\Gamma \nu _1 $ and $ \nu
_{0,2}=(1-\beta \cos \theta )\Gamma \nu _2$.

\subsection{Formulas in terms of the integral of $\cos \protect\theta $}

In terms of the integral of $\cos \theta $, the forms of the above
formulas become simpler. Let
\begin{equation}
\mu \equiv \cos \theta ,
\end{equation}
\begin{equation}
\mu _{\max }\equiv \cos \widetilde{\theta }_{\min }=\min \{\cos
\theta _{\min },\frac{(t_{0,\max }-t_{0,c})\Gamma +\frac
Dc-(t-t_c)}{(t_{0,\max }-t_{0,c})\Gamma \beta +\frac{R_c}c}\}
\end{equation}
and
\begin{equation}
\mu _{\min }\equiv \cos \widetilde{\theta }_{\max }=\max \{\cos
\theta _{\max },\frac{(t_{0,\min }-t_{0,c})\Gamma +\frac
Dc-(t-t_c)}{(t_{0,\min }-t_{0,c})\Gamma \beta +\frac{R_c}c}\},
\end{equation}
where (13)-(14) are applied.

The flux now can be calculated by
\begin{equation}
f_\nu (t)=\frac{2\pi [(t-t_c-\frac Dc)\beta c+R_c]^2}{D^2\Gamma ^3}%
\int_{\mu _{\min }}^{\mu _{\max }}\frac{I_{0,\nu }(t_0,\nu _0)\mu
}{(1-\beta \mu )^5}d\mu ,
\end{equation}
where $t_0$ and $\mu $ are related by
\begin{equation}
t_0=\frac{t-t_c-\frac Dc+\frac{R_c}c\mu }{(1-\beta \mu )\Gamma
}+t_{0,c}
\end{equation}
and $\nu _0$ and $\mu $ are related by $ \nu _0=(1-\beta \mu )\Gamma
\nu $.

The count rate is now determined by
\begin{equation}
\frac{dn(t)}{dt}=\frac{2\pi [(t-t_c-\frac Dc)\beta c+R_c]^2}{D^2\Gamma ^3h%
}\int_{\mu _{\min }}^{\mu _{\max }}[\int_{\nu _{0,1}}^{\nu _{0,2}}\frac{%
I_{0,\nu }(t_0,\nu _0)}{\nu _0}d\nu _0]\frac \mu {(1-\beta \mu
)^5}d\mu ,
\end{equation}
where $ \nu _{0,1}=(1-\beta \mu )\Gamma \nu _1$ and $ \nu
_{0,2}=(1-\beta \mu )\Gamma \nu _2$.

\subsection{Formulas in terms of the integral of the proper emission time}

Since the rest frame intensity $I_{0,\nu }(t_0,\nu _0)$ is always
provided in the form of the function of the proper emission time
$t_0$, it would be more convenient to compute the flux or the count
rate with formulas in terms of the integral of $t_0$.

From (24) we get
\begin{equation}
\mu =\frac{(t_0-t_{0,c})\Gamma +\frac Dc-(t-t_c)}{\frac{R_c}%
c+(t_0-t_{0,c})\Gamma \beta },
\end{equation}
and from  $ \nu _0=(1-\beta \mu )\Gamma \nu $ we find
\begin{equation}
\nu _0=\frac{\frac{R_c}c-[\frac Dc-(t-t_c)]\beta }{ \frac{R_c}%
c+(t_0-t_{0,c})\Gamma \beta }\Gamma \nu .
\end{equation}
According to (26) and (27) one can check that $\frac{dt_0}{d\mu
}>0$.

Replacing $\mu $ with $t_0$, we obtain the following form of formula
for calculating the flux
\begin{equation}
f_\nu (t)=\frac{2\pi c^2\int_{\widetilde{t}_{0,\min }}^{\widetilde{t}%
_{0,\max }}I_{0,\nu }(t_0,\nu _0)[(t_0-t_{0,c})\Gamma +\frac Dc-(t-t_c)][%
\frac{R_c}c+(t_0-t_{0,c})\Gamma \beta ]^2dt_0}{D^2\Gamma ^2\{\frac{R_c}%
c-[\frac Dc-(t-t_c)]\beta \}^2},
\end{equation}
where $\widetilde{t}_{0,\min }$ and $\widetilde{t}_{0,\max }$ are
determined by
\begin{equation}
\widetilde{t}_{0,\min }=\max \{t_{0,\min },\frac{t-t_c-\frac Dc+\frac{R_c}%
c\cos \theta _{\max }}{(1-\beta \cos \theta _{\max })\Gamma
}+t_{0,c}\}
\end{equation}
and
\begin{equation}
\widetilde{t}_{0,\max }=\min \{t_{0,\max },\frac{t-t_c-\frac Dc+\frac{R_c}%
c\cos \theta _{\min }}{(1-\beta \cos \theta _{\min })\Gamma
}+t_{0,c}\},
\end{equation}
respectively, and $\nu _0$ and $t_0$ are related by (27).

The count rate is now written in the form
\begin{equation}
\frac{dn(t)}{dt}=\frac{2\pi c^2\int_{\widetilde{t}_{0,\min }}^{\widetilde{%
t}_{0,\max }}[\int_{\nu _1}^{\nu _2}I_{0,\nu }(t_0,\nu _0)\frac{d\nu
}\nu ][(t_0-t_{0,c})\Gamma +\frac
Dc-(t-t_c)][\frac{R_c}c+(t_0-t_{0,c})\Gamma \beta ]^2dt_0}{D^2\Gamma
^2h\{\frac{R_c}c-[\frac Dc-(t-t_c)]\beta \}^2}.
\end{equation}
According to (27), in the case that $t$ and $t_0$ could be treated
as constants we get $ \frac{d\nu }\nu
|_{t_0=const,t=const}=\frac{d\nu _0}{\nu _0}$. Replacing variable
$\nu $ with $\nu _0$ we get from (31) that
\begin{equation}
\frac{dn(t)}{dt}=\frac{2\pi c^2\int_{\widetilde{t}_{0,\min }}^{\widetilde{%
t}_{0,\max }}[\int_{\nu _{0,1}}^{\nu _{0,2}}\frac{I_{0,\nu }(t_0,\nu _0)}{%
\nu _0}d\nu _0][(t_0-t_{0,c})\Gamma +\frac Dc-(t-t_c)][\frac{R_c}%
c+(t_0-t_{0,c})\Gamma \beta ]^2dt_0}{D^2\Gamma
^2h\{\frac{R_c}c-[\frac Dc-(t-t_c)]\beta \}^2},
\end{equation}
where $\nu _{0,1}=[\frac{R_c}c-(\frac Dc-t+t_c)\beta ]\Gamma \nu _1
/ [\frac{R_c} c+(t_0-t_{0,c})\Gamma \beta ] $ and $\nu
_{0,2}=[\frac{R_c}c-(\frac Dc-t+t_c)\beta ]\Gamma \nu _2 /
[\frac{R_c} c+(t_0-t_{0,c})\Gamma \beta ] $.

\subsection{Formulas in terms of the integral of relative timescales}

Assign [19]
\begin{equation}
\tau ^{\prime }\equiv \frac{t^{\prime }-t_c}{\frac{R_c}c}, \qquad
\tau _{\min }^{\prime }\equiv \frac{t_{\min }^{\prime
}-t_c}{\frac{R_c}c}, \qquad \tau _{\max }^{\prime }\equiv
\frac{t_{\max }^{\prime }-t_c}{\frac{R_c}c},
\end{equation}%
\begin{equation}
\tau _0\equiv \frac{t_0-t_{0,c}}{\frac{R_c}c}, \qquad \tau _{0,\min
}\equiv \frac{t_{0,\min }-t_{0,c}}{\frac{R_c}c}, \qquad \tau
_{0,\max }\equiv \frac{t_{0,\max }-t_{0,c}}{\frac{R_c}c},
\end{equation}
and
\begin{equation}
\tau \equiv \frac{t-t_c-\frac Dc+\frac{R_c}c}{\frac{R_c}c}.
\end{equation}
One finds that
\begin{equation}
\nu _0=\frac{1-\beta +\beta \tau }{1+\beta \Gamma \tau _0}\Gamma \nu
,
\end{equation}
\begin{equation}
\tau ^{\prime }=\frac{\tau -(1-\cos \theta )}{1-\beta \cos \theta },
\end{equation}
\begin{equation}
0\leq \tau _{\min }^{\prime }\leq \tau ^{\prime }\leq \tau _{\max
}^{\prime },
\end{equation}
and
\begin{equation}
1-\cos \theta _{\min }+(1-\beta \cos \theta _{\min })\tau _{\min
}^{\prime }\leq \tau \leq 1-\cos \theta _{\max }+(1-\beta \cos
\theta _{\max })\tau _{\max }^{\prime }
\end{equation}
[which is the range of $\tau $ within which the radiation defined
within (3) and (4) is observable]. In addition, one gets $ \tau
_{0,\min }=\frac{\tau _{\min }^{\prime }}\Gamma $, $ \tau _{0,\max
}=\frac{\tau _{\max }^{\prime }}\Gamma $, and
\begin{equation}
1-\cos \theta _{\min }+(1-\beta \cos \theta _{\min })\Gamma \tau
_{0,\min }\leq \tau \leq 1-\cos \theta _{\max }+(1-\beta \cos \theta
_{\max })\Gamma \tau _{0,\max }.
\end{equation}

Let
\begin{equation}
\widetilde{I}_{0,\nu }(\tau _0,\nu _0)\equiv I_{0,\nu }[t_0(\tau
_0),\nu _0].
\end{equation}

We get from (28) that
\begin{equation}
f_\nu (\tau )=\frac{2\pi R_c^2\int_{\widetilde{%
\tau }_{0,\min }}^{\widetilde{\tau }_{0,\max }}\widetilde{I}_{0,\nu
}(\tau
_0,\nu _0)(1-\tau +\tau _0\Gamma )(1+\Gamma \beta \tau _0)^2d\tau _0}{%
D^2\Gamma ^2(1-\beta +\beta \tau )^2},
\end{equation}
where $\nu _0$ is related to $\tau $ and $\nu $ by (36),
\begin{equation}
\widetilde{\tau }_{0,\min }=\frac{\widetilde{t}_{0,\min }-t_{0,c}}{\frac{%
R_c}c}=\max \{\tau _{0,\min },\frac{\tau -1+\cos \theta _{\max
}}{(1-\beta \cos \theta _{\max })\Gamma }\},
\end{equation}
and
\begin{equation}
\widetilde{\tau }_{0,\max }=\frac{\widetilde{t}_{0,\max }-t_{0,c}}{\frac{%
R_c}c}=\min \{\tau _{0,\max },\frac{\tau -1+\cos \theta _{\min
}}{(1-\beta \cos \theta _{\min })\Gamma }\}.
\end{equation}

Let
\begin{equation}
C(\tau )\equiv \frac{dn[t(\tau )]}{d\tau }.
\end{equation}
We then get
\begin{equation}
C(\tau )=\frac{2\pi R_c^3\int_{\widetilde{\tau }_{0,\min }}^{\widetilde{%
\tau }_{0,\max }}[\int_{\nu _{0,1}}^{\nu
_{0,2}}\frac{\widetilde{I}_{0,\nu }(\tau _0,\nu _0)}{\nu _0}d\nu
_0](1-\tau +\Gamma \tau _0)(1+\beta \Gamma \tau _0)^2d\tau
_0}{hcD^2\Gamma ^2(1-\beta +\beta \tau )^2}.
\end{equation}
where $
\nu _{0,1}=\frac{(1-\beta +\beta \tau )\Gamma }{1+\beta \Gamma \tau _0}%
\nu _1$ and $
\nu _{0,2}=\frac{(1-\beta +\beta \tau )\Gamma }{1+\beta \Gamma \tau _0}%
\nu _2$. Note that (42) and (46) hold only when (40) is satisfied.
In the range beyond (40), $f_\nu (\tau )$ and $C(\tau )$ will become
zero.

Formula (46) shows that the profile of count rates of a fireball
source is a function of $\tau $. It is independent of the real time
scale $t$, and independent of the real size, $R_c$, of the source.
In other words, no matter how large is the fireball concerned and
how large is the observed timescale concerned, for the profile of
the count rate, only the ratio of the latter to the time scale of
the initial radius of the fireball plays a role.

Applying (33) and (34) to  $ t^{\prime }-t_c=\Gamma (t_0-t_{0,c})$
we get $ \tau ^{\prime }=\Gamma \tau _0$. Substituting it into (37)
comes to
\begin{equation}
\tau _0=\frac{\tau -(1-\cos \theta )}{(1-\beta \cos \theta )\Gamma
}.
\end{equation}
Generally, observation time $\tau $ is related to the co-moving time
$\tau _0 $ of the surface with $\theta $ by
\begin{equation}
\tau =(1-\beta \cos \theta )\Gamma \tau _0+1-\cos \theta .
\end{equation}
For the same emission time $\tau _0$, photons emitted from surfaces
with different line-of-sight angles would reach the observer at
different times. In units of the time scale of the initial radius
$R_c$, the time delay of
photons emitted from $\theta $ relative to that of those emitted from $%
\theta =0$ is $1-\cos \theta $. The contraction factor of the
emission time is $(1-\beta \cos \theta )\Gamma $ which differs from
surface to surface. Taking $\theta =0$ we gain
\begin{equation}
\tau =(1-\beta )\Gamma \tau _0\qquad \qquad (\theta =0).
\end{equation}
This is the relation that relates the observation time of photons
which are emitted from $\theta =0$ with the corresponding co-moving
emission time. The time is contracted by a factor of $(1-\beta
)\Gamma $.

\section{Spectral evolution within the period of fireball pulses coming from a typical intrinsic emission form}

As revealed in [9], the observed hardness ratio is seen to be harder
at the beginning of some GRB pulses than what the pure curvature
effect (when no intrinsic spectral evolution is involved) could
predict and be softer at later times of the pulses, causing a
so-called ``harder-leading'' problem. An economic mechanism was
suggested to account for this problem, which is the hard-to-soft
evolution pattern of the rest frame spectrum of fireballs (see [9]).
In a more realistic situation, before the hardest spectrum appears,
there might be a short period within which the co-moving spectrum
undergoes a soft-to-hard phase. This would happen when the inner
shell possesses a relatively harder core and surrounding the core
are attached with less dense materials. The hardest spectrum will
appear when the core strikes the outer shell. Before that the
spectrum radiated from the outer shell is softer since the shell is
hit by relatively less dense materials and then gains a much smaller
acceleration and a smaller speed.

Let us investigate the spectral evolution in a simple case where the
emission of a co-moving pulse with an exponential rise and
exponential decay and with a flexible Comptonized radiation form
from the whole fireball surface is observed. For the sake of
comparison, we consider three co-moving pulses (or rest frame
pulses) with their spectra being unchanged, hard-to-soft, and
soft-to-hard-to-soft respectively during the period concerned. The
co-moving pulse with an unchanged spectrum is assumed to be
\begin{equation}
\widetilde{I}_{0,\nu }(\tau _0,\nu _0)=I_0\nu _0^{1+\alpha _C}\exp
(-\nu _0/\nu _{0,C})\{
\begin{array}{c}
\exp (-\frac{\tau _{0,0}-\tau _0}{\sigma _r})\qquad (\tau _{0,\min
}\leq
\tau _0\leq \tau _{0,0}) \\
\exp (-\frac{\tau _0-\tau _{0,0}}{\sigma _d})\qquad \qquad \qquad
(\tau
_{0,0}<\tau _0)%
\end{array}
.
\end{equation}
The co-moving pulse (50) is intrinsically the same as that adopted
in [9]. [Note that, $ \tau ^{\prime }=\Gamma \tau _0$, or $ \tau
_0\propto \tau ^{\prime }$.] The co-moving pulse possessing a
hard-to-soft spectral evolution is assumed to be
\begin{equation}
\widetilde{I}_{0,\nu }(\tau _0,\nu _0)=I_0\nu _0^{1+\alpha _C}\exp
[-(\tau _0/\tau _{0,0})(\nu _0/\nu _{0,C})]\{
\begin{array}{c}
\exp (-\frac{\tau _{0,0}-\tau _0}{\sigma _r})\qquad (\tau _{0,\min
}\leq
\tau _0\leq \tau _{0,0}) \\
\exp (-\frac{\tau _0-\tau _{0,0}}{\sigma _d})\qquad \qquad \qquad
(\tau
_{0,0}<\tau _0)%
\end{array}
,
\end{equation}
where $\tau _{0,0}\neq 0$. The co-moving pulse with a
soft-to-hard-to-soft spectrum is assumed to be
\begin{equation}
\widetilde{I}_{0,\nu }(\tau _0,\nu _0)=I_0\nu _0^{1+\alpha _C}\{
\begin{array}{c}
\exp [-(\tau _{0,0}/\tau _0)(\nu _0/\nu _{0,C})]\exp (-\frac{\tau
_{0,0}-\tau _0}{\sigma _r})\qquad (\tau _{0,\min }<\tau _0\leq \tau
_{0,0})
\\
\exp [-(\tau _0/\tau _{0,0})(\nu _0/\nu _{0,C})]\exp (-\frac{\tau
_0-\tau
_{0,0}}{\sigma _d})\qquad \qquad \qquad (\tau _{0,0}<\tau _0)%
\end{array}
,
\end{equation}
where $\tau _0>0$ is required (as shown below, this will easily be
satisfied). As adopted previously, we assign $\tau _0=10\sigma
_r+\tau _{0,\min }$. According to (4), we assign $t^{\prime }\geq
t_c$, which leads to $\tau _{0,\min }\geq 0$. As long as this
condition is satisfied, $\tau _{0,\min }$ could be freely chosen.
[Note that, in the case that $t^{\prime }\geq t_c$ does not hold,
$\tau _{0,\min }$ is also constrained: $\tau _{0,\min }>-(\beta
\Gamma )^{-1}$, when $\beta
>0$ (see [19]). Without any loss of generality we take $\tau
_{0,\min }=0$. To meet the condition that the soft-to-hard period is
much shorter than the hard-to-soft phase in the third co-moving
pulse, we take $\sigma _d=2\sigma _r$ and this will be applied to
all the three co-moving pulses in the following analysis.

\begin{figure}[tbp]
\begin{center}
\includegraphics[width=5in,angle=0]{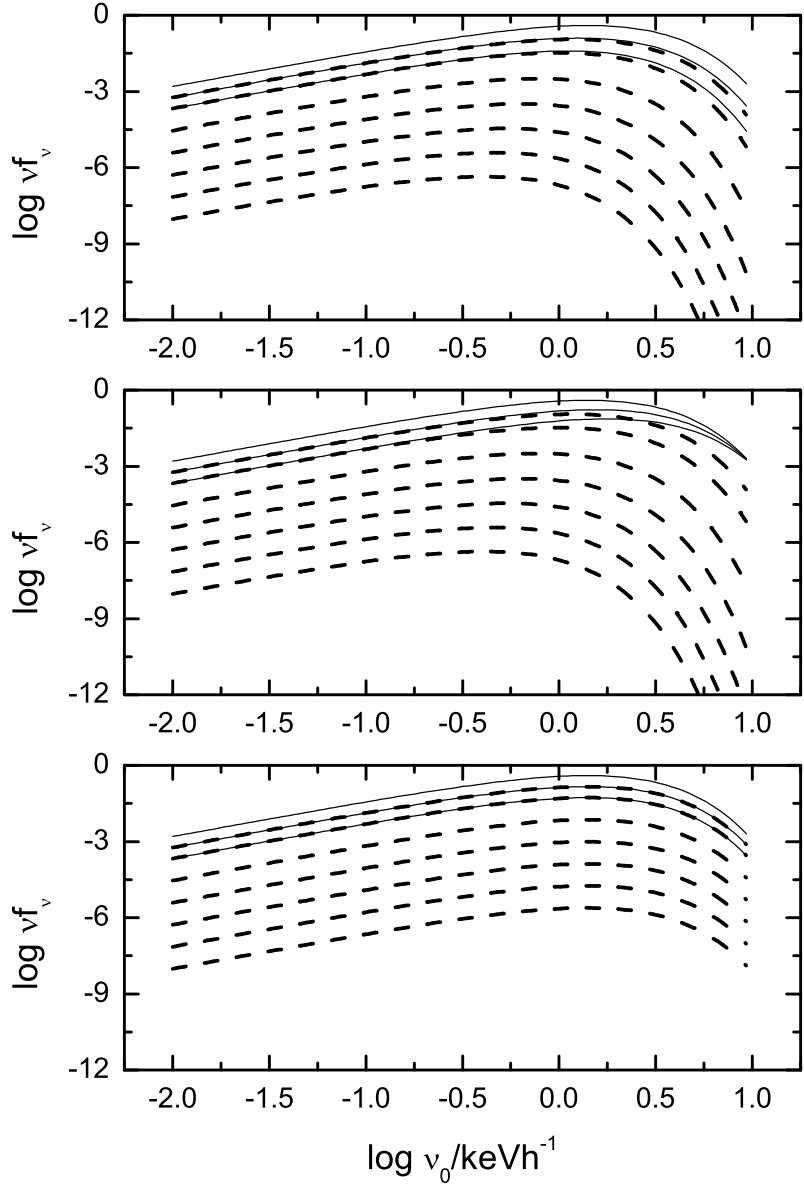}
\end{center}
\caption{Spectral evolutions of co-moving pulses (50) (the lower
panel), (51) (the mid panel), and (52) (the upper panel), where we
take $ \sigma _r=0.01$, $ \sigma _d=2\sigma _r$, $ \alpha
_{0,C}=-0.6$, $ \nu _{0,C} = 1 keV h^{-1}$, and $I_0=1$. Solid lines
from the bottom to the top stand for the co-moving times $ \tau
_{0}(1)$, $ \tau _{0}(2)$, and $ \tau _{0}(3)$, respectively, and
dashed lines from the top to the bottom represent $ \tau _{0}(4)$, $
\tau _{0}(5)$, $ \tau _{0}(6)$, $ \tau _{0}(7)$, $ \tau _{0}(8)$, $
\tau _{0}(9)$ , and $ \tau _{0}(10)$, respectively.} \label{Fig. 1}
\end{figure}

Shown in Fig. 1 are the spectral evolutions of the three co-moving
pulses (50), (51), and (52), where the co-moving times concerned are
$\tau _{0}(1)=\tau _{0,0}-2\sigma _r$, $\tau _{0}(2)=\tau
_{0,0}-\sigma _r$, $\tau _{0}(3)=\tau _{0,0}$, $\tau _{0}(4)=\tau
_{0,0}+\sigma _d$, $\tau
_{0}(5)=\tau _{0,0}+2\sigma _d$, $\tau _{0}(6)=\tau _{0,0}+4\sigma _d$, $%
\tau _{0}(7)=\tau _{0,0}+6\sigma _d$, $\tau _{0}(8)=\tau
_{0,0}+8\sigma _d$, $\tau _{0}(9)=\tau _{0,0}+10\sigma _d$, and
$\tau _{0}(10)=\tau _{0,0}+12\sigma _d$, respectively.

\begin{figure}[tbp]
\begin{center}
\includegraphics[width=4in,angle=0]{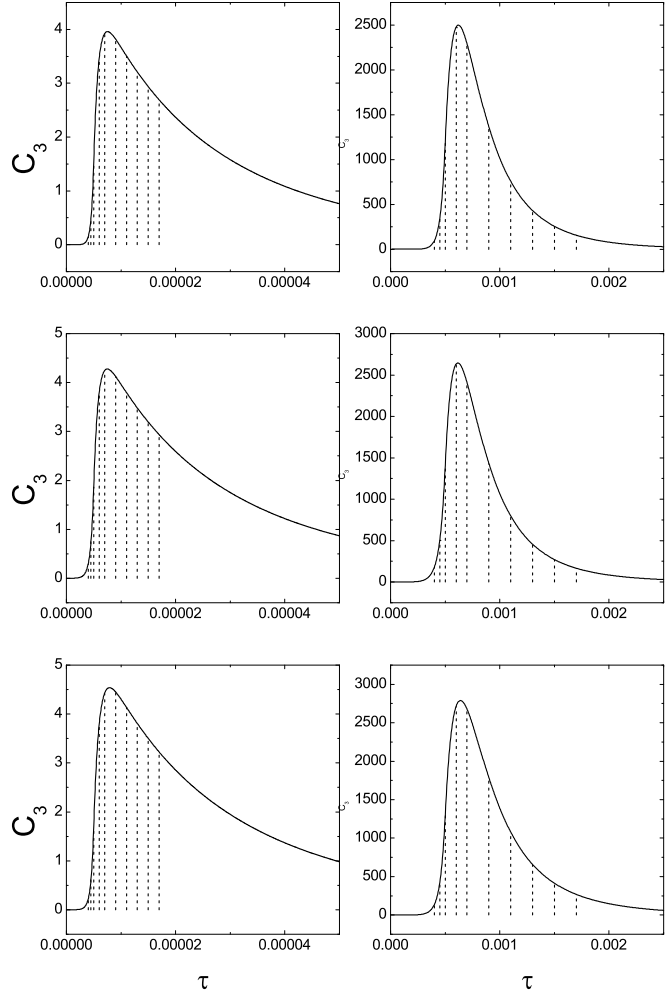}
\end{center}
\caption{Positions of the observation times corresponding to various
co-moving times for photons emitted from the tip of the fireball ($
\theta =0$) marked in the light curves expected by distant
observers. The light curves associated with co-moving pulses (50)
(lower panels), (51) (mid panels), and (52) (upper panels) are
calculated with equation (46), where we take $ \sigma _d=2 \sigma
_r$, $ \alpha _{0,C}=-0.6$, $ \nu _{0,C} = 1 keV h^{-1}$,
$\Gamma=100$, and $C_0\equiv2 \pi R_c^3I_0/hcD^2=1$. The width in
the rising part of the co-moving pulses is taken as $ \sigma
_r=0.0001$ (left panels) and $ \sigma _r=0.01 $ (right panels),
which correspond to ``narrow'' and ``broad'' pulses, respectively
(see [17] and [19]). Solid lines stand for the light curves expected
in the third channel of BATSE ($55keV\leq E<110keV$). Dashed lines
from the left to the right represent observation times $ \tau (1)$,
$ \tau (2)$, $ \tau (3)$, $ \tau (4)$, $ \tau (5)$, $ \tau (6)$, $
\tau (7)$, $ \tau (8)$, $ \tau (9)$, and $ \tau (10)$,
respectively.} \label{Fig. 2}
\end{figure}

Let us check how these co-moving pulses give rise to the spectra of
fireballs. According to (49), the observation times of photons which
are emitted from the tip of the fireball (i.e., $\theta =0$) at the
above co-moving times are $\tau (1)=(1-\beta )\Gamma(\tau
_{0,0}-2\sigma _r)$, $\tau (2)=(1-\beta )\Gamma(\tau _{0,0}-\sigma
_r)$, $\tau (3)=(1-\beta )\Gamma(\tau _{0,0})$, $ \tau (4)=(1-\beta
)\Gamma(\tau _{0,0}+\sigma _d)$, $\tau (5)=(1-\beta )\Gamma(\tau
_{0,0}+2\sigma _d)$, $\tau (6)=(1-\beta )\Gamma(\tau _{0,0}+4\sigma
_d)$, $\tau (7)=(1-\beta )\Gamma(\tau _{0,0}+6\sigma _d)$, $ \tau
(8)=(1-\beta )\Gamma(\tau _{0,0}+8\sigma _d)$, $\tau (9)=(1-\beta
)\Gamma(\tau _{0,0}+10\sigma _d)$, and $\tau (10)=(1-\beta
)\Gamma(\tau _{0,0}+12\sigma _d)$, respectively. Positions of these
times in the light curves observed by distant observers are shown in
Fig. 2, where, (46) is employed to calculate the corresponding light
curves. Note that, equation (21) in [19] could only be applied to a
simple case when the time component and the spectral component of an
intrinsic pulse could be separated (see equation (10) in [19]). The
advantage of equation (46) is that it could be directly applied to a
co-moving radiation which can take any forms. For the same reason,
(42) will be applied to calculate the observed spectrum in the
following analysis.

In Fig. 2, two typical widthes of co-moving pulses are adopted to
show how the co-moving width influences the observed spectrum. They
are $\sigma _r=0.0001$ and $\sigma _r=0.01 $ in co-moving pulses
(50)-(52), which correspond to relatively narrow and broad pulses
respectively (see [17] and [19]). The words ``narrow'' and ``broad''
refer to the profile of the observed pulses. A ``narrow'' pulse
comes from a local pulse with a small ratio of its width to the
radius of the fireball [or, the ratio of the co-moving pulse width
to the product of the radius of the fireball and the Lorentz factor
is small; see equation (49)]. A character of this kind of pulse is
that the pulse possesses a more steadily decaying phase. Or more
precisely, the deviation of its decaying profile to the so-called
standard decaying form is mild (for a detailed explanation, see
[17]). Conversely, a ``broad'' pulse arises from a local pulse with
a large ratio of its width to the radius of the fireball. A
character of this kind of pulse is that its decaying profile
possesses a reverse S-feature deviation from the standard decaying
curve (or the deviation is very obvious) [17].

\subsection{Evolutionary pattern of the overall spectrum}

Displayed in Fig. 3 are the spectral evolutions within the period of
the expected observed pulses of fireballs, which are outcomes of the
three co-moving pulses (50), (51), and (52). We use equation (42) to
calculate the spectra measured at the observation times $\tau (1)$,
$\tau (2)$, $\tau (3)$, $\tau (4)$, $\tau (5)$, $\tau (6)$, $\tau
(7)$, $\tau (8)$, $\tau (9)$, and $\tau (10)$. The resulted overall
spectra show a general hard-to-soft evolution pattern within the
period of the pulses observed. During these observation times, the
spectral evolution of ``narrow'' pulses is mild (see Fig. 3 left
panels), while for ``broad'' pulses their spectra develop quite
rapidly (see Fig. 3 right panels). In the decaying part of light
curves of ``narrow'' or ``broad'' pulses, the spectral evolution
patterns are quite similar for the three co-moving pulses (see
dashed lines in Fig. 3). This must be due to the fact that the
decaying part of the pulses is dominated by high latitude emission
of the fireball, where the curvature effect is important. It is then
natural that the spectral evolutions show a similar trend. In the
rising part of the light curves of pulses, the situation is
different. The spectral evolution pattern of co-moving pulses
obviously influences the way the observed spectrum evolves.

\begin{figure}[tbp]
\begin{center}
\includegraphics[width=4in,angle=0]{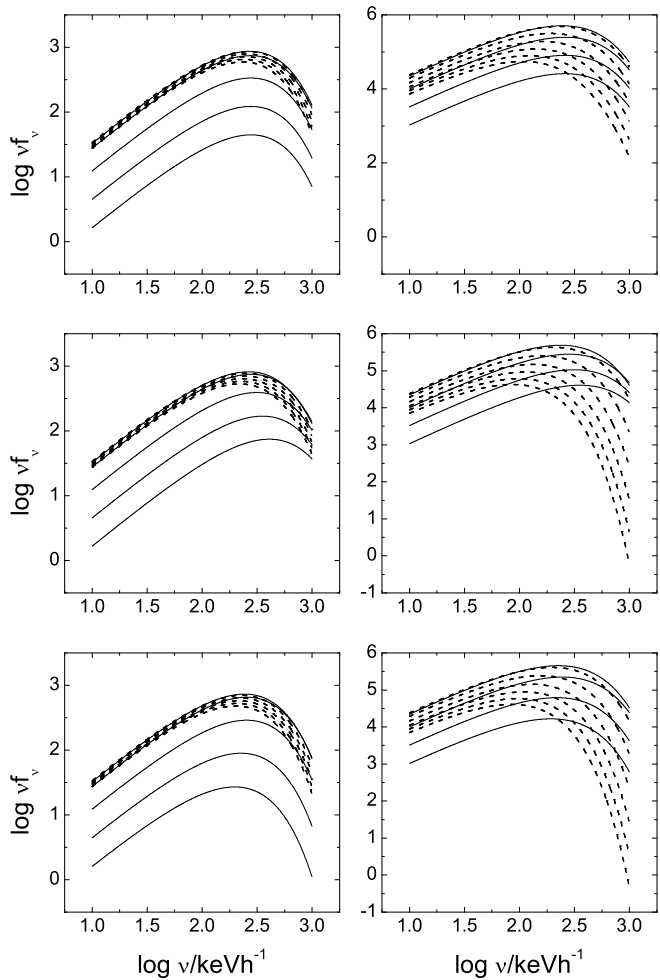}
\end{center}
\caption{Spectral evolutions of fireball pulses arising from
co-moving pulses (50) (lower panels), (51) (mid panels), and (52)
(upper panels). The spectra are calculated with (42). Each panel in
this plot corresponds to the same panel in Fig. 2, where the
parameters are the same. The only difference is that we take
$F_0\equiv2 \pi R_c^2I_0/D^2=1$. In the left panels, solid lines
from the bottom to the top stand for the spectra measured at
observation times $ \tau (1)$, $ \tau (2)$, $ \tau (3)$, $ \tau
(4)$, and $ \tau (5)$, respectively, while dashed lines from the top
to the bottom represent the spectra measured at $ \tau (6)$, $ \tau
(7)$, $ \tau (8)$, $ \tau (9)$, and $ \tau (10)$, respectively. In
the right panels, solid lines from the bottom to the top stand for
the spectra measured at $ \tau (1)$, $ \tau (2)$, $ \tau (3)$, and $
\tau (4)$, respectively, and dashed lines from the top to the bottom
denote the spectra measured at $ \tau (5)$, $ \tau (6)$, $ \tau
(7)$, $ \tau (8)$, $ \tau (9)$, and $ \tau (10)$, respectively.}
\label{Fig. 3}
\end{figure}

\subsection{Evolution of the peak energy of the $\nu f_{\nu}$ spectrum}

There are two quantities capable of describing the hardness of
spectra. One is the so-called hardness ratio and the other is the
peak energy $E_p$, the energy where the peak of $\nu f_{\nu}$ is
observed. Evolutions of the two quantities could describe in some
extent how the spectrum evolves. Recently, the hardness ratio curves
associated with different situations were discussed [9]. Here, we
pay our attention mainly to the evolution of the peak energy within
the period of fireball pulses which arise from the three co-moving
pulses proposed above.

We employ the following integrated flux (see Section 5) to compare
the evolution of $E_p$ with light curves:
\begin{equation}
F\equiv \int_{\nu _{1}}^{\nu _{2}}f_{\nu }d\nu
\end{equation}
where $\nu _{1}$ and $\nu _{2}$ are the lower and upper limits of
the energy channel concerned.

\begin{figure}[tbp]
\begin{center}
\includegraphics[width=4.5in,angle=0]{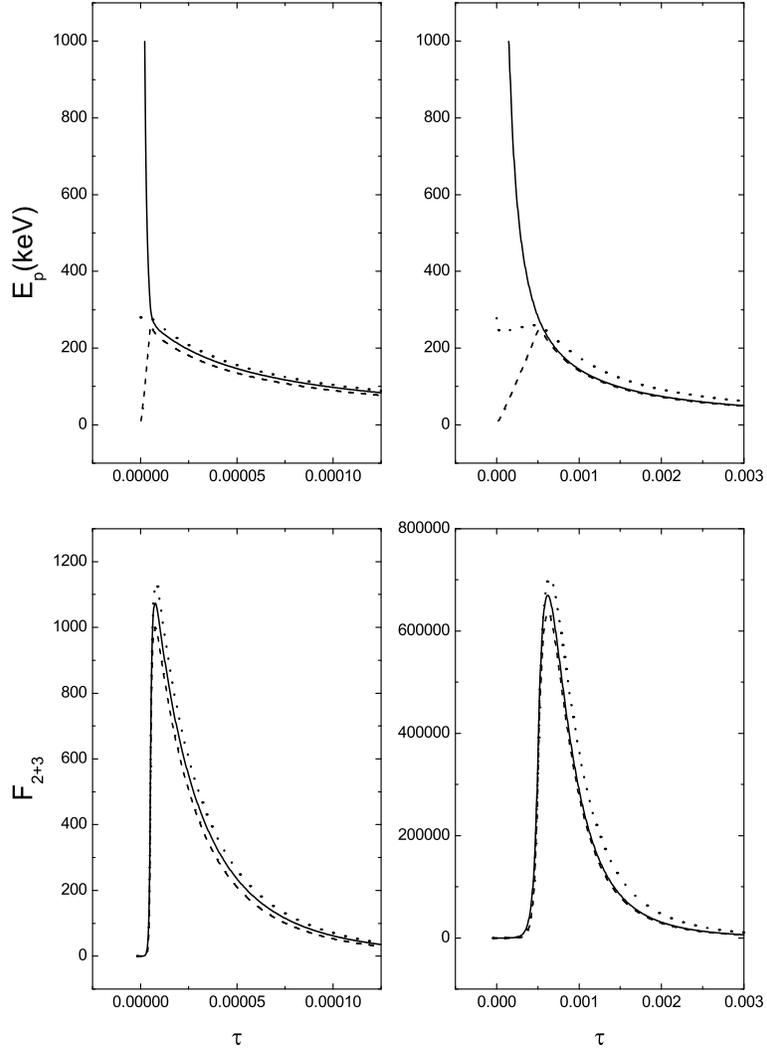}
\end{center}
\caption{Evolutionary curves of the peak energy (upper panels) of
fireball pulses arising from co-moving pulses (50) (dotted lines),
(51) (solid lines), and (52) (dashed lines). For the sake of
comparison, we present in the lower panels the integrated flux
$F_{2+3}$ over the the energy range of the second ($55keV\leq
E<110keV$) and third ($110keV\leq E<320keV$) BATSE channels, which
are calculated with (42) and (53). Parameters are the same as they
are in Fig. 2. As adopted in Fig. 2, the width in the rising part of
the co-moving pulses is taken as $ \sigma _r=0.0001$ (left panels)
and $ \sigma _r=0.01 $ (right panels), which correspond to
``narrow'' and ``broad'' pulses, respectively.} \label{Fig. 4}
\end{figure}

Displayed in Fig. 4 are the developments of the peak energy in
various cases considered above. We find that the feature of
drop-to-rise-to-drop revealed in the hardness ratio curve in [9] is
observed in the peak energy evolutionary curve and it holds only for
the unchanged intrinsic spectrum. In the case of intrinsically
hard-to-soft spectrum, we observe a continuous hard-to-soft pattern,
while in the case of intrinsically soft-to-hard-to-soft spectrum,
one finds a soft-to-hard-to-soft evolution pattern for the observed
spectrum. There is a turning point in the $E_p(t)$ curve (hereafter
EPC). After the turning point, the EPCs associated with the three
co-moving pulses show the same trend of softening in the case of
``narrow'' pulses. Before the turning point, the observed spectra
vary significantly in accordant with their intrinsic spectral
evolution patterns. This suggests that the spectrum observed before
the turning point must mainly be determined by the intrinsic
evolution pattern, while after the turning point, it is dominated by
the curvature effect. In the ``broad'' pulse cases, the softening of
the spectra arising from the un-changed intrinsic spectrum co-moving
pulses are different from that of the hard-to-soft and
soft-to-hard-to-soft intrinsic spectrum co-moving pulses, which
leads to a noticeable character discussed below. The sharpness of
the turning feature might become an indicator showing a pulse
observed is ``narrow'' or ``broad''.

\subsection{Comparison of the spectrum of the emission of the tip
with that arising from the overall fireball surface}

If the fireball moves relativistically towards the observer but not
expands, one would observed an entirely different evolution pattern
of the spectrum within the pulse concerned. Compared with this
spectrum, is that arising from the expanding fireball harder or
softer? An investigation on this issue might bring us some
interesting information. The spectrum of an ejecta moving towards
the observer must be the same as that emitted from the tip of an
expanding fireball surface when the intrinsic spectra and the
Lorentz factors are the same.

The spectrum of the emission from the tip of the fireball surface
must merely be a blue-shifting of the intrinsic one. Due to the
contribution of high latitude emissions, a deviation of the observed
spectrum of the whole fireball from that of the tip is expected. If
the deviation is small within some period of observation time, then
one can estimates the form of the intrinsic spectrum from the
observed one within this time interval, although the blue-shifting
factor and the amplifying factor (or, the boosting factor) remain
unknown. Let us investigate how this deviation evolves and what it
depends on.

\subsubsection{In the corresponding overall spectra}

The method for comparing the expected tip spectrum with that of the
emission from the whole fireball surface is straightforward. One can
simply shift the co-moving spectra shown in Fig. 1 to high energy
range where the emission of the tip is expected, and then compares
them with those presented in Fig. 3. According to the Doppler
effect, photons emitted from the tip with frequency $\nu_0$ would be
blue shifted to $\nu = \nu_0/\Gamma (1-\beta)$ when they reach the
observer. We will use this shifting factor, $1/\Gamma (1-\beta)$, to
shift the spectra in Fig. 1.

\begin{figure}[tbp]
\begin{center}
\includegraphics[width=4in,angle=0]{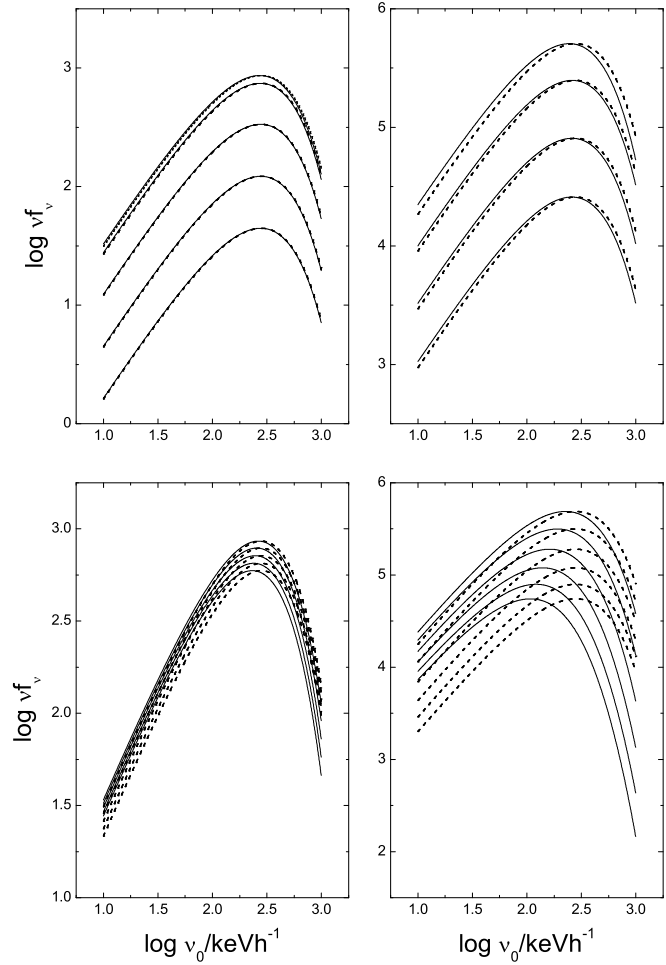}
\end{center}
\caption{Comparison between spectra expected from the emission of
the tip (dashed lines) and the emission from the whole fireball
surface (solid lines) in cases of ``narrow'' (left panels) and
``broad'' (right panels) pulses arising from the co-moving pulse
with an unchanged intrinsic spectrum, that of (50). Solid lines in
the left upper and lower panels are the solid and dashed lines in
the left lower panel of Fig. 3, respectively. Solid lines in the
right upper and lower panels are the solid and dashed lines in the
right lower panel of Fig. 3, respectively. Dashed lines denote the
shifting lines of the corresponding co-moving spectra presented in
the lower panel of Fig. 1, which correspond to the co-moving times
(see the caption of Fig. 1) when photons emitted at these times from
the tip of the fireball surface reach the observer at the times
concerned in Fig. 3 (see the captions of Figs. 2 and 3).}
\label{Fig. 5}
\end{figure}

\begin{figure}[tbp]
\begin{center}
\includegraphics[width=4in,angle=0]{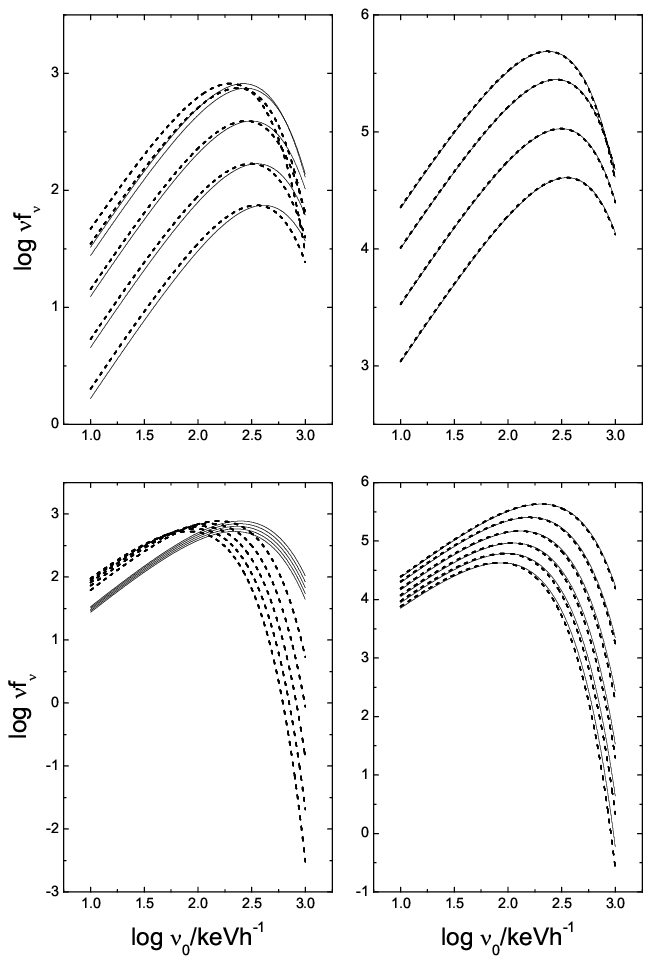}
\end{center}
\caption{Comparison between spectra expected from the emission of
the tip (dashed lines) and the emission from the whole fireball
surface (solid lines) in cases of ``narrow'' (left panels) and
``broad'' (right panels) pulses arising from the co-moving pulse
with a hard-to-soft intrinsic spectrum, that of (51). Solid lines in
the left upper and lower panels are the solid and dashed lines in
the left mid panel of Fig. 3, respectively. Solid lines in the right
upper and lower panels are the solid and dashed lines in the right
mid panel of Fig. 3, respectively. Dashed lines denote the shifting
lines of the corresponding co-moving spectra presented in the mid
panel of Fig. 1. (See Fig. 3 for a more detailed explanation.)}
\label{Fig. 6}
\end{figure}

\begin{figure}[tbp]
\begin{center}
\includegraphics[width=4in,angle=0]{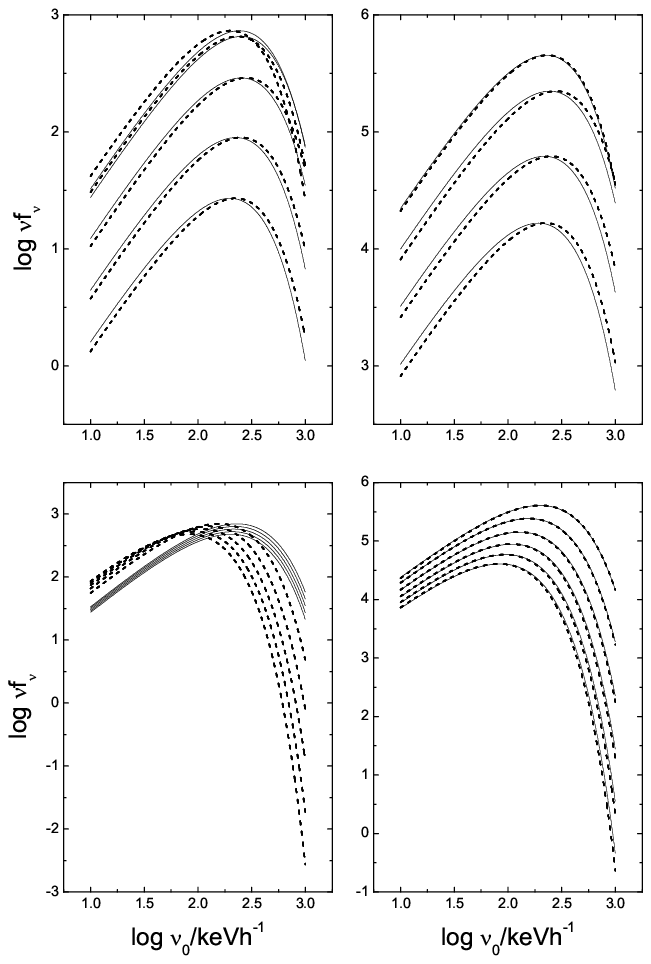}
\end{center}
\caption{Comparison between spectra expected from the emission of
the tip (dashed lines) and the emission from the whole fireball
surface (solid lines) in cases of ``narrow'' (left panels) and
``broad'' (right panels) pulses arising from the co-moving pulse
with a soft-to-hard-to-soft intrinsic spectrum, that of (52). Solid
lines in the left upper and lower panels are the solid and dashed
lines in the left upper panel of Fig. 3, respectively. Solid lines
in the right upper and lower panels are the solid and dashed lines
in the right upper panel of Fig. 3, respectively. Dashed lines
denote the shifting lines of the corresponding co-moving spectra
presented in the upper panel of Fig. 1. (See Fig. 3 for a more
detailed explanation.)} \label{Fig. 7}
\end{figure}

Shown in Figs. 5, 6 and 7 are the comparisons between spectra
expected from the emission of the tip (which is merely the shifting
of the intrinsic spectrum) and the emission from the whole fireball
surface, made for co-moving pulses (50), (51), and (52),
respectively. For the sake of comparison, we normalize the shifting
spectra in Fig. 1 to the maximums of the corresponding spectra in
Fig. 3.

In the case of the unchanged intrinsic spectrum, the observed
spectrum in the rising part of the light curve for ``narrow'' pulses
is almost the same as that expected from the emission of the tip of
fireballs (see the left upper panel of Fig. 5), while it deviates
slightly from the latter for ``broad'' pulses (see the right upper
panel of the figure). In the decaying portion, the observed spectra
are obviously softer than that expected from the emission of the tip
of fireballs, and the softening is very significant for ``broad''
pulses (see the two lower panels of Fig. 5). This indicates that if
an unchanged spectrum during the rising part of pulses is observed,
then the corresponding intrinsic spectrum is likely an unchanged
one.

When the evolution of the intrinsic spectrum is hard-to-soft, the
observed spectrum is hard-to-soft as well. In the rising part of the
light curve of ``broad'' pulses, the spectrum is almost the same as
that of the emission expected from the tip of fireballs (see the
left upper panel of Fig. 6), while in the decaying portion of the
light curve it deviates slightly from the latter (see the left lower
panel of the figure). For ``narrow'' pulses, the spectrum is always
harder than that of the emission expected from the tip of fireballs,
and the deviation is obvious and it is very significant in the
decaying portion of the light curve.

In the case of the soft-to-hard-to-soft intrinsic spectrum, the
observed spectra in the rising part of the light curve are softer
than, but quite close to, that expected from the emission of the tip
of fireballs. In the decaying phase, they are much harder than the
latter for ``narrow'' pulses and are almost the same as the latter
for ``broad'' pulses.

\subsubsection{In terms of the peak energy}

One might notice that the magnitude of the spectrum of the emission
expected from the tip of fireballs discussed about is not real since
emission from other parts of fireballs is much larger than that
merely from the tip and hence they are not comparable. However, the
concept of this spectrum is useful since it does represent the real
form of emission from the tip. After all, it is the form that
determines how ``hard'' is a spectrum.

Let us investigate how the spectrum arising from the whole fireball
surface deviates from that of the emission of the tip of fireballs
in terms of the peak energy. We define the deviation of the peak
energy of the spectrum of the whole fireball emission and that of
the tip as:
\begin{equation}
\Delta E_p \equiv E_p - E_{p,tip},
\end{equation}
where $E_{p,tip}$ is the peak energy of the spectrum of the emission
expected from the tip.

\begin{figure}[tbp]
\begin{center}
\includegraphics[width=5in,angle=0]{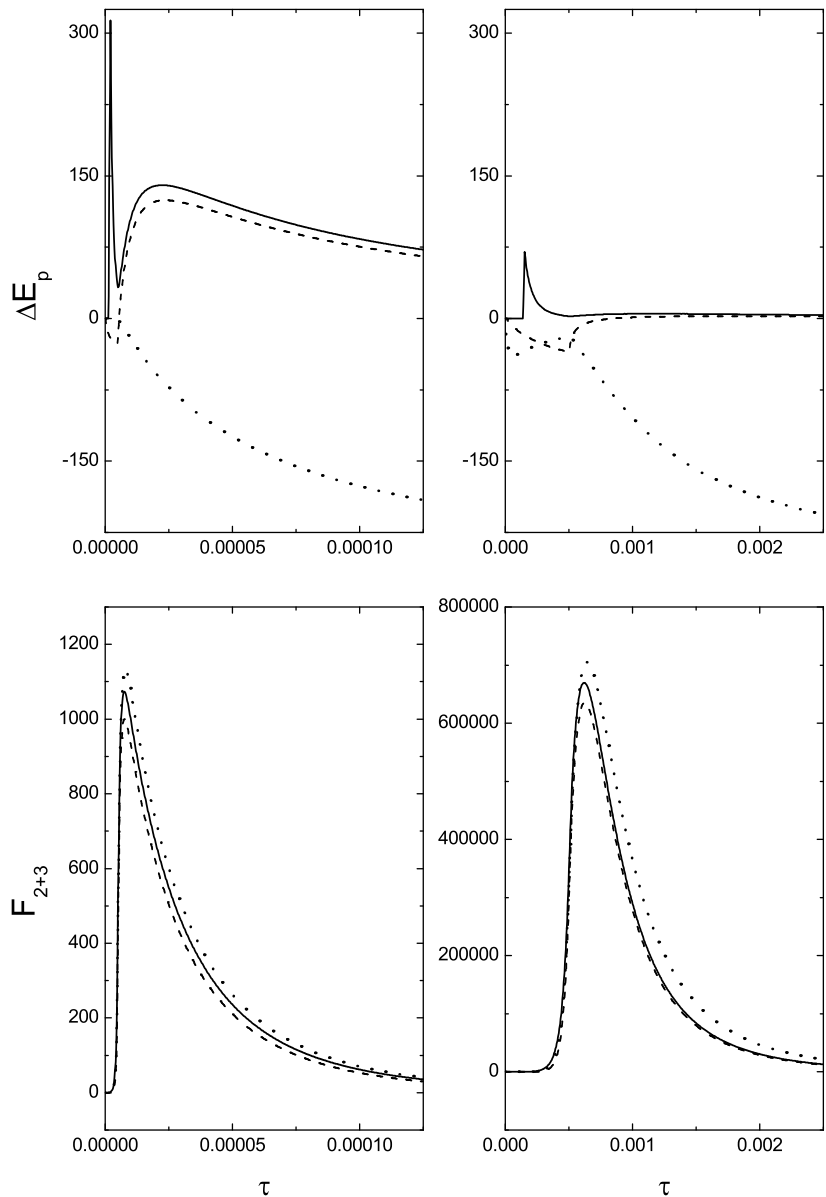}
\end{center}
\caption{Evolutions of $\Delta E_p$ (upper panels) for ``narrow''
(left panels) and ``broad'' (right panels) pulses, calculated with
(54). Light curves of the integrated flux $F_{2+3}$ (lower panels)
are also presented (see Fig. 4) for a direct comparison. Parameters
and symbols are the same as they are in Fig. 4.} \label{Fig. 8}
\end{figure}

Displayed in Fig. 8 are the relations between $\Delta E_p$ and time
derived from various cases. Conclusions obtained above are
reinforced. a) In the case of the unchanged intrinsic spectrum, that
arising from the whole fireball surface is always softer than the
spectrum merely coming from the tip (see also Fig. 5). In the
decaying phase of the pulses, the softening becomes stronger and
stronger. b) When the intrinsic spectrum evolves in a simple
hard-to-soft pattern, the spectrum from the whole fireball surface
is always harder than that merely coming from the tip (which is
beyond our initial expectation) (see also Fig. 6). c) For a
soft-to-hard-to-soft spectral evolution, the spectrum of the whole
fireball surface becomes softer than that from the tip during the
rising phase of the pulses, and then turns to be harder than the
latter in the decaying phase (see also Fig. 7). d) For ``broad''
pulses, the hardening in the decaying phase in the latter two cases
is very mild and thus the observed spectrum could serve as a good
estimator of that from the tip emission. e) In the case of
``narrow'' pulses, the hardening in the decaying phase is so strong
that one cannot estimate the intrinsic spectrum from an observed
one. A consequence of this phenomenon is that in the decaying phase,
``narrow'' pulses are generally harder than ``broad'' ones. This
could possibly be checked by current observation.

One might notice that the deviation of the peak energy is sensitive
in the rising part of pulses. In the case of the simple hard-to-soft
intrinsic spectral evolution pattern, the observed peak energy could
become about $300keV$ larger than that expected from the tip of
fireballs when the pulses observed are relatively narrow, and it
could be about $70keV$ larger than that of the tip when the pulses
concerned are relatively broad. In the decaying phase, $E_p$ could
be about $50-150keV$ larger than that of the tip for ``narrow''
pulses whilst it is almost the same as that of the tip for ``broad''
ones.

\subsection{Spectral evolution revealed in other aspects}

\subsubsection{Relation between the flux and peak energy}

An important aspect capable of revealing spectral evolution in GRB
pulses is the well-known relation between the flux and peak energy.
Revealed in Fig. 2 of [29], the observed flux $F$ was found to be
significantly correlated with the peak energy $E_p$ within a single
burst for a subset of their sample. In their study, $F$ is the
integrated flux in a wide energy range which was also adopted in
[30]. Here we explore the relation between the integrated flux and
peak energy expected from a fireball pulse. Several cases discussed
above are studied on this issue.

\begin{figure}[tbp]
\begin{center}
\includegraphics[width=5in,angle=0]{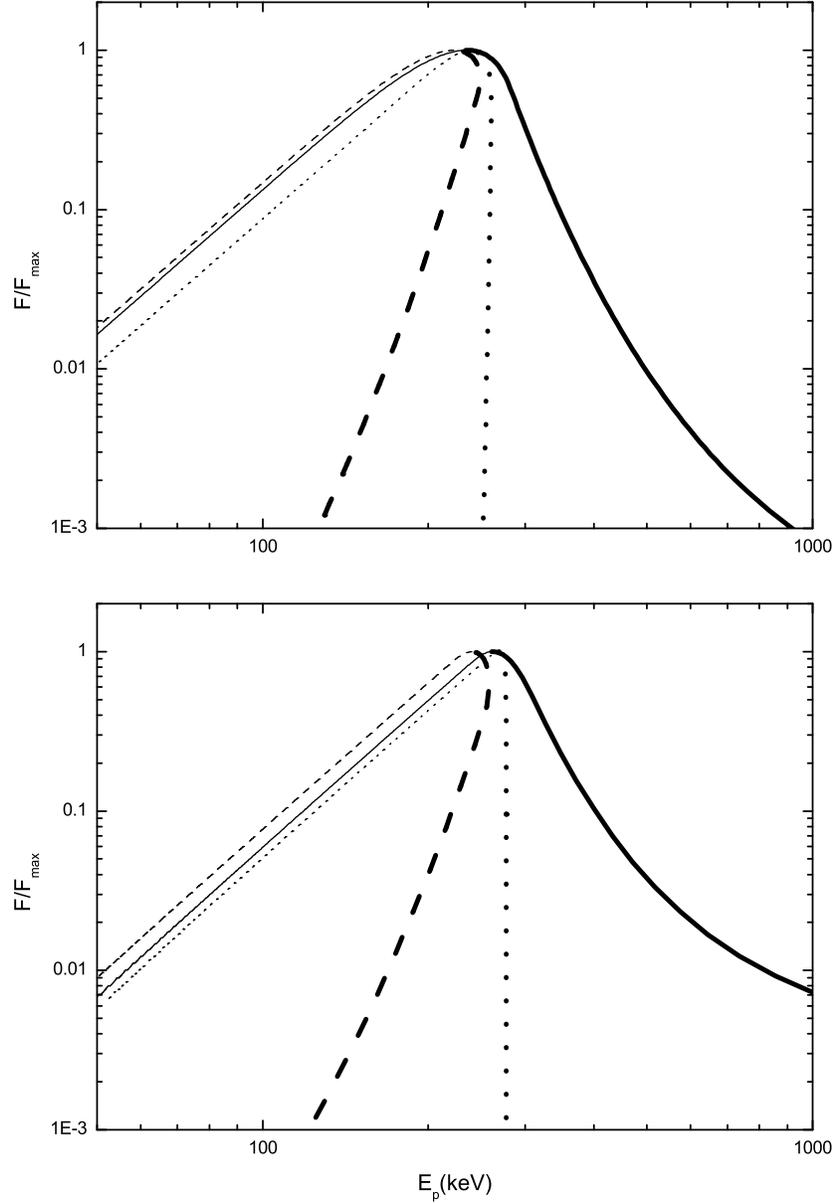}
\end{center}
\caption{Relations between the integrated flux and peak energy in
cases of the unchanged intrinsic spectrum (dotted lines),
hard-to-soft intrinsic spectrum (solid lines), and
soft-to-hard-to-soft intrinsic spectrum (dashed lines) for
``narrow'' (the lower panel) and ``broad'' (the upper panel) pulses
of fireballs. Thin lines denote the relation in the decaying phase
of pulses and thick lines stand for that in the rising part of the
light curve of pulses.} \label{Fig. 9}
\end{figure}

The results are shown in Fig. 9, where the adopted energy range
confining the integrated flux is $(1,10000)keV$. A linear relation
between the logarithm of the two quantities is seen in the decaying
phase of pulses. In terms of statistics, the relation is called the
hardness-intensity correlation (HIC) which was noticed by many
authors in gamma-ray bursts [24], [28-29], [31-32]. There is a
turnover feature in the relation between the two quantities. The
turnover feature shows a hook-like curve which is sensitive to
intrinsic spectral evolution patterns. An intrinsic
soft-to-hard-to-soft spectral evolution corresponds to a semi-linear
correlation between $log F$ and $log E_p$ in the rising phase of
pulses, with its slope being larger than that in the decaying phase.
In the case of an unchanged intrinsic spectrum, the relation in the
rising phase is a straight line in parallel with the $F$ axis. When
the intrinsic spectral evolution pattern is a simple hard-to-soft
one, the observed integrated flux $F$ would decrease with the
increasing of $E_p$ in the rising phase.

Relations between the integrated flux and peak energy in the
decaying phase of all the six pulses discussed above are calculated.
They are found to strictly follow a power law: $ F\propto
E_{p}^{\gamma}$, with $\gamma\sim 3$. This is in agreement with what
was found by [16] in a recent investigation.

\subsubsection{Relevant relations}

The first relevant relation discussed is that between the peak value
of the $\nu f_{\nu}$ spectrum and the peak energy $E_p$, which was
previously investigated by [41]. We study all the six pulses
discussed above and find that the relation is similar to that
between the flux and peak energy (the figure is very similar to Fig.
9, which is omitted). We find the same power law relation $ (\nu
f_{\nu})_p\propto E_{p}^{\gamma}$ in the decaying phase of pulses
and approximately the same index $\gamma\sim 3$.

The second is the relation between the flux $f_{\nu}$ at a fixed
energy $E$ and the spectral peak energy $E_p$. Here we adopt
$E=200keV$. All the six pulses discussed above are studied. With a
slight difference, the relation is similar to that between the flux
and peak energy (the figure is omitted). Although $f_{\nu}$
increases with the increasing of $E_p$ in the decaying phase of
pulses, the relation is no more a power law.

\begin{figure}[tbp]
\begin{center}
\includegraphics[width=5in,angle=0]{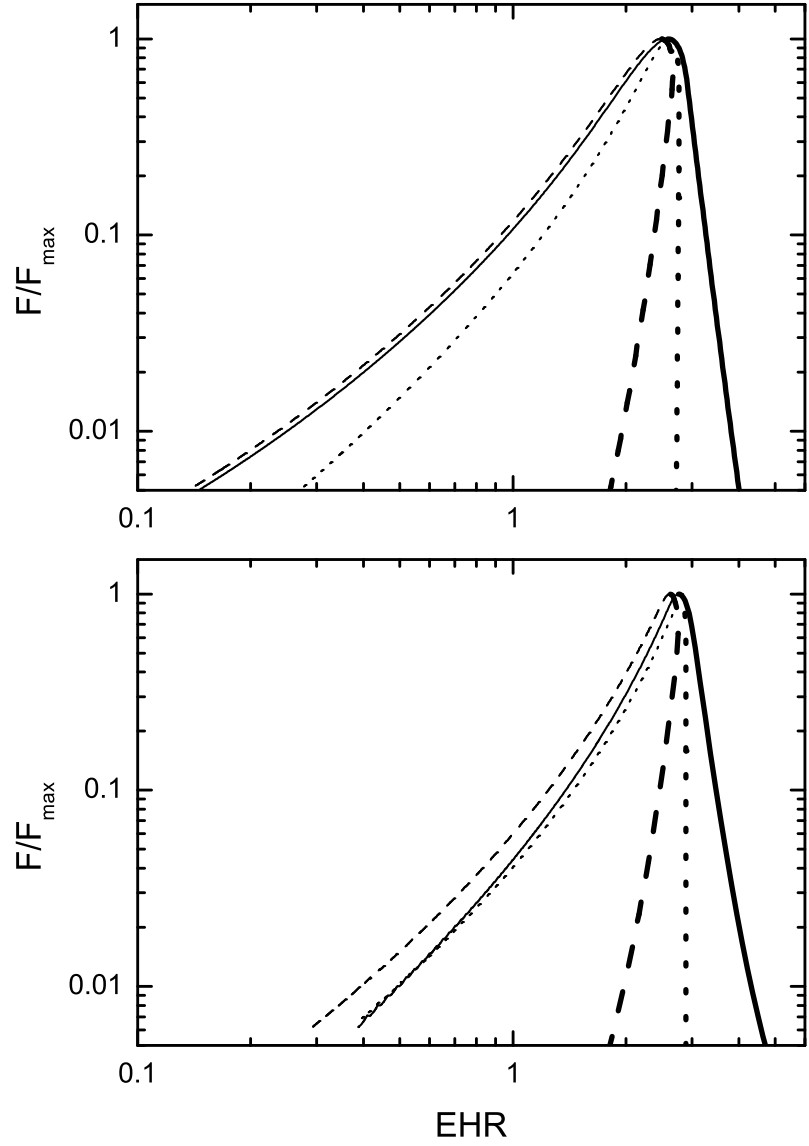}
\end{center}
\caption{Relations between the integrated flux and the hardness
ratio. The symbols are the same as they are in Fig. 9.} \label{Fig.
13}
\end{figure}

The third is the relation between the integrated flux over the
energy range adopted above (see Fig. 9) and the hardness ratio. The
hardness ratio adopted here is defined by $EHR\equiv F_3/F_2$, where
$F_2$ and $F_3$ are the integrated fluxes confined in the energy
ranges of the second and third BATSE channels, respectively. Shown
in Fig. 10 is the relation in various cases discussed in Fig. 9. In
the decaying phase of the six pulses concerned, the relation is not
a power law, but still, the integrated flux increases with the
increasing of the hardness ratio.

\section{Spectral evolution of pulses associated with other intrinsic emission forms}

The real GRB emission might be more complicated than what discussed
above. We wonder if different intrinsic emission forms could lead to
other conclusions.

Following [16], we adopt an intrinsic broken-power-law (BPL)
spectrum in the emission of a pulse. This pulse is assumed to
possess a linear rise and linear decay phases. In the same way we
consider three evolutionary patterns which are unchanged,
hard-to-soft, and soft-to-hard-to-soft respectively.

The co-moving pulse is assumed to be
\begin{equation}
\widetilde{I}_{0,\nu }(\tau _0,\nu _0)=I_0\{
\begin{array}{c}
(\frac{\nu _0}{\nu _{0,b}})^{1+\alpha} \qquad (\nu _0\leq \nu _{0,b}) \\
(\frac{\nu _0}{\nu _{0,b}})^{1+\beta} \qquad (\nu _{0,b}<\nu _0)
\end{array}
\}\{
\begin{array}{c}
\frac{\tau _0-\tau _{0,\min }}{\tau _{0,0}-\tau _{0,\min }}\qquad
(\tau
_{0,\min }<\tau _0\leq \tau _{0,0}) \\
\frac{\tau _{0,\max }-\tau _0}{\tau _{0,\max }-\tau _{0,0}}\qquad
(\tau _{0,0}<\tau _0< \tau _{0,\max })
\end{array}
,
\end{equation}
where $\nu _{0,b}$ is a function of $\tau _0$. Co-moving pulse (81)
is a pulse with a linear rise and a linear decay, emitting with a
BPL radiation form. In the case of an unchanged intrinsic spectrum,
$\nu _{0,b}$ is taken as $ \nu _{0,b}=\nu _{0,0}$ when $\tau
_{0,min}<\tau _0< \tau _{0,\max }$ (where $\nu _{0,0}$ is a
constant). In the case of a hard-to-soft intrinsic spectrum, $\nu
_{0,b}$ is assumed to decrease linearly with the increasing of $\tau
_0$: $ \nu _{0,b}=\frac{\tau _{0,0}-\tau _{0,\min }}{\tau _0-\tau
_{0,\min }}\nu _{0,0}$ when $\tau _{0,min}<\tau _0< \tau _{0,\max
}$. In the case of a soft-to-hard-to-soft intrinsic spectrum, $\nu
_{0,b}$ is assumed to rise linearly and then drop also linearly with
the increasing of $\tau _0$: $\nu _{0,b}=\frac{\tau _0-\tau _{0,\min
}}{\tau _{0,0}-\tau _{0,\min }}\nu _{0,0}$ when $\tau _{0,\min
}<\tau _0\leq \tau _{0,0}$, $\nu _{0,b}=\frac{\tau _{0,\max }-\tau
_0}{\tau _{0,\max }-\tau _{0,0}}\nu _{0,0}$ when $\tau _{0,0}<\tau
_0< \tau _{0,\max }$.

One might notice that the broken-power-law model could well
approximate the Band function model [33] in very low and high energy
ranges. We accordingly assign the typical values of the low and high
energy indexes available from the fit of the Band function model to
BATSE GRB spectra to the BPL model used here: $\alpha=-1$ and
$\beta=-2.25$ [34-35]. In addition, we assign $\tau _{0,\min }=0$
and take $\tau _{0,\max }-\tau _{0,0}=2(\tau _{0,0}-\tau _{0,\min
})$. We take $\Gamma =100$ so that a direct comparison with those
results in Section 3 could be made.

We study the evolutionary curve of the peak energy (EPC) in the case
of the broken power law emission (55) emitted with unchanged,
hard-to-soft, and soft-to-hard-to-soft intrinsic spectra for
``narrow'' and ``broad'' pulses of fireballs (the half width in the
rising phase of the ``narrow'' co-moving pulse is taken as 0.0001,
and that of the ``broad'' one is 0.01). The analysis shows that
conclusions drawn from Fig. 4 hold in this situation (the figure is
omitted). Noticeable differences are: in the case of ``narrow''
pulses, the EPC arising from the hard-to-soft intrinsic spectrum is
significantly larger than that arising from the unchanged and
soft-to-hard-to-soft intrinsic spectra in the decaying phase; in the
case of ``broad'' pulses, the EPC arising from the unchanged
intrinsic spectrum is larger than that arising from the other two
intrinsic spectra in the decaying phase and it seems to deviate from
the common trend observed in other cases.

\begin{figure}[tbp]
\begin{center}
\includegraphics[width=5in,angle=0]{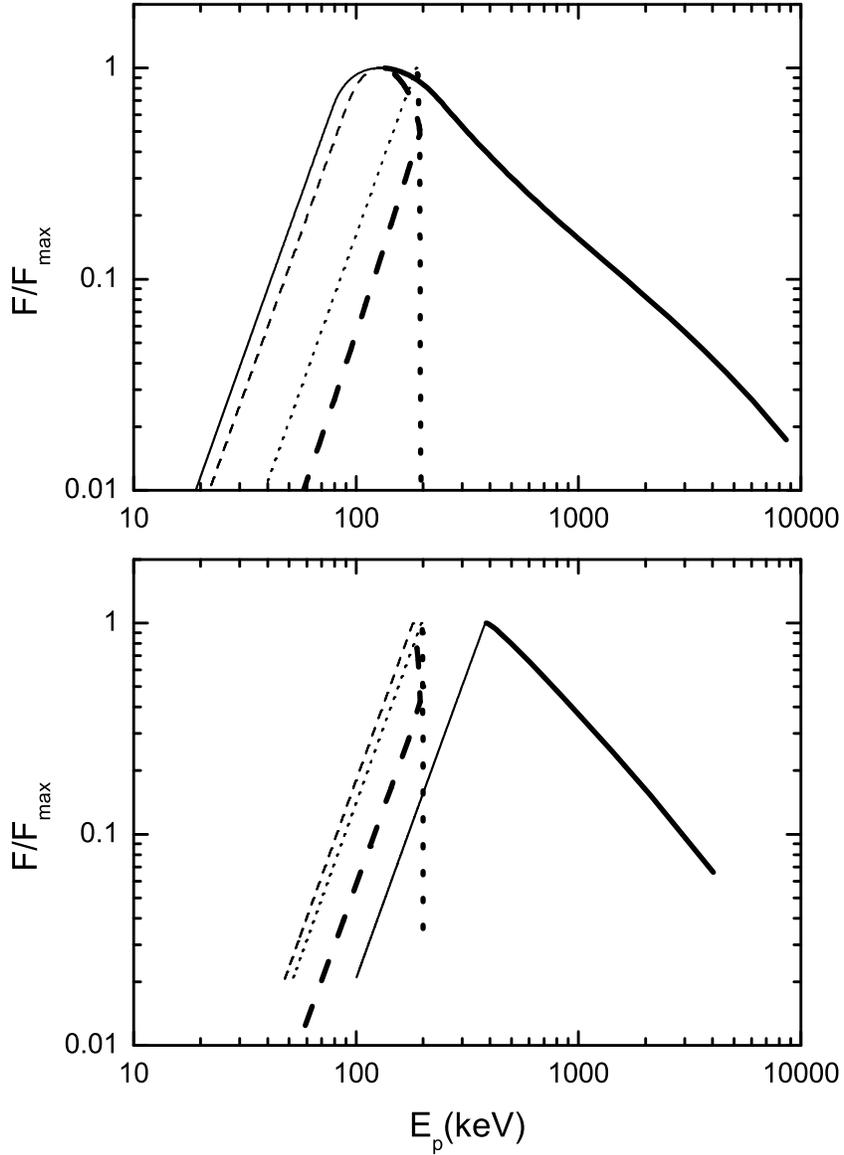}
\end{center}
\caption{Relations between the integrated flux and peak energy in
the case of the broken power law emission (55) with an unchanged
intrinsic spectrum (dotted lines), hard-to-soft intrinsic spectrum
(solid lines), and soft-to-hard-to-soft intrinsic spectrum (dashed
lines) for ``narrow'' (the lower panel) and ``broad'' (the upper
panel) pulses of fireballs. Parameters or their relations are:
$\alpha=-1$, $\beta=-2.25$, $\tau _{0,\min }=0$, $\tau _{0,\max
}-\tau _{0,0}=2(\tau _{0,0}-\tau _{0,\min })$, and $\Gamma =100$.
The half width in the rising phase of the ``narrow'' co-moving pulse
is taken as 0.0001, and that of the ``broad'' one is 0.01. The thick
and thin lines stand for the same as they do in Fig. 9.} \label{Fig.
15}
\end{figure}

Relations between the integrated flux and peak energy for the three
intrinsic pulses confined by the three $\nu _{0,b} (\tau _0)$
functions mentioned above are shown in Fig. 11. The turnover
features shown in Fig. 9 are observed in this figure. The power law
relation between the two quantities in the decaying phase of pulses
holds for all cases concerned here, where the index is found to be
$\gamma\sim 3$ as well. In addition, we notice that the whole curves
of the $F - E_p$ relation vary significantly with the intrinsic
emission, and the ranges of the power law relation depends strongly
on the emission as well.

\section{Spectral evolution of pulses expected in other situations of the soft-to-hard-to-soft intrinsic emission}

In the analysis of the soft-to-hard-to-soft intrinsic emission of
co-moving pulses performed above, the peak energy of the intrinsic
emission concerned is that starting from and ending at very low
energy and the hardest intrinsic spectrum occurs at the time when
the peak flux of the co-moving pulse appears. This might not be true
in practice. In fact, a co-moving pulse might start to emit at a
high energy band and then evolve to the maximum and then drop to the
minimum which still remains in a high energy band. In particular,
some pulses might arise from the emission that the hardest intrinsic
spectrum appears ahead of the peak flux. We are anxious if
conclusions obtained above are affected when taking all these into
account.

Two co-moving pulses with a skew soft-to-hard-to-soft spectrum,
starting to emit at a high energy band and ending its emission at a
relatively lower band are considered. One is a modified Comptonized
radiation form (52):
\begin{equation}
\widetilde{I}_{0,\nu }(\tau _0,\nu _0)=I_0\nu _0^{1+\alpha _C}\{
\begin{array}{c}
0\qquad \qquad \qquad \qquad \qquad \qquad \qquad \qquad \qquad
(\tau
_{0,\min }<\tau _0\leq \tau _{0,1}) \\
\exp [-(\tau _{0,p}/\tau _0)(\nu _0/\nu _{0,C})]\exp (-\frac{\tau
_{0,0}-\tau _0}{\sigma _r})\qquad (\tau _{0,1}<\tau _0\leq \tau _{0,p}) \\
\exp [-(\tau _0/\tau _{0,p})(\nu _0/\nu _{0,C})]\exp (-\frac{\tau
_{0,0}-\tau _0}{\sigma _r})\qquad (\tau _{0,p}<\tau _0\leq \tau _{0,0}) \\
\exp [-(\tau _0/\tau _{0,p})(\nu _0/\nu _{0,C})]\exp (-\frac{\tau
_0-\tau _{0,0}}{\sigma _d})\qquad (\tau _{0,0}<\tau _0\leq \tau
_{0,2})
\end{array}
,
\end{equation}
where $\tau _{0,p}=\tau _{0,0}-\Delta \tau _0$ and $\Delta \tau _0$
represents the offset of the time the hardest spectrum appears ahead
of that of the peak flux. We assign $\tau _{0,1}=\tau _{0,p}/2$ and
$\tau _{0,2}=2\tau _{0,p}$. In applying co-moving pulse (56), we
take $\Delta \tau_0=\sigma _r$, $\sigma _r=0.01$ and $\Gamma=100$,
and adopt all other parameters as the same of those adopted in
Section 3 (see the caption of Fig. 1).

The other is a revised broken-power-law emission form which is still
described by (55), with $\nu _{0,b}$ being determined by: $\nu
_{0,b}=\frac{\tau _0+\tau _{0,p}-2\tau _{0,\min }}{2(\tau
_{0,p}-\tau _{0,\min })}\nu _{0,0}$ when $\tau _{0,\min }<\tau
_0\leq \tau _{0,p}$, $\nu _{0,b}=\frac{2\tau _{0,\max }-\tau
_{0,p}-\tau _0}{2(\tau _{0,\max }-\tau _{0,p})}\nu _{0,0}$ when
$\tau _{0,p}<\tau _0<\tau _{0,\max }$ (where $\tau _{0,p}=\tau
_{0,0}-\Delta \tau _0$). In applying co-moving pulse (55) coupling
this relation, we assign $\tau _{0,\min }=0$ and take $\tau _{0,\max
}-\tau _{0,0}=2(\tau _{0,0}-\tau _{0,\min })$, $\alpha=-1$,
$\beta=-2.25$, and $\Gamma =100$, as adopted in last section. We
consider only a ``broad'' co-moving pulse and therefore take its
half width in the rising phase as 0.01.

\begin{figure}[tbp]
\begin{center}
\includegraphics[width=5in,angle=0]{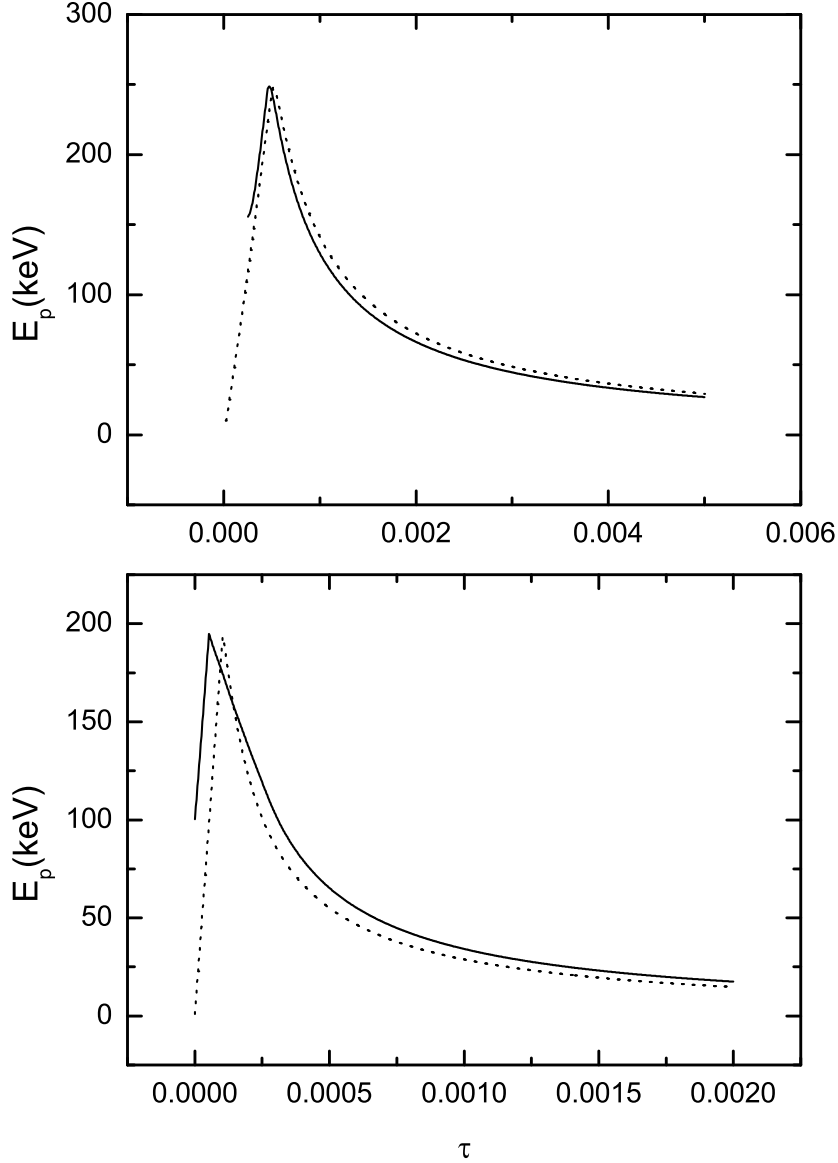}
\end{center}
\caption{Evolutionary curves of the peak energy associated with
co-moving pulses with a skew soft-to-hard-to-soft spectrum. In the
upper panel, the solid line is the curve arising from co-moving
pulse (56); the dotted line is the dashed line in the upper left
panel of Fig. 4. In the lower panel, the solid line is the curve
arising from co-moving pulse (55) coupling the $\nu _{0,b} (\tau
_0)$ function presented in Section 5; the dotted line  is the curve
arising from co-moving pulse (55) coupling the soft-to-hard-to-soft
$\nu _{0,b} (\tau _0)$ function presented in Section 4. The pulses
concerned are ``broad'' ones (see Section 5 for the details of the
adopted parameters).} \label{Fig. 17}
\end{figure}

\begin{figure}[tbp]
\begin{center}
\includegraphics[width=5in,angle=0]{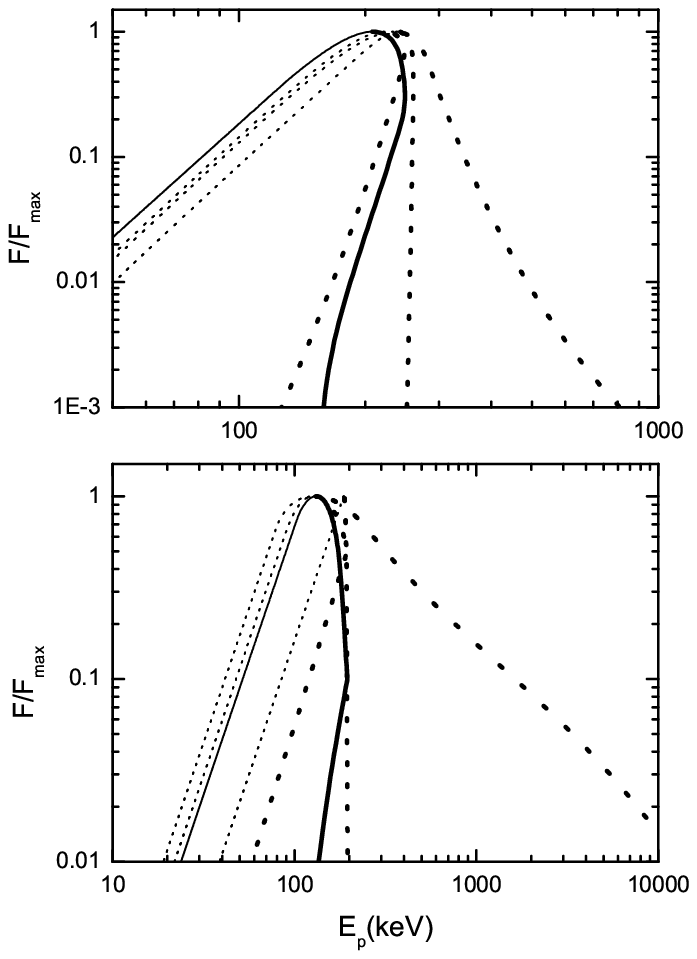}
\end{center}
\caption{Relation between the integrated flux and peak energy
associated with co-moving pulses with a skew soft-to-hard-to-soft
spectrum. In the upper panel, the solid line is the curve arising
from co-moving pulse (56); the dotted lines represent all the lines
in the upper panel of Fig. 9. In the lower panel, the solid line is
the curve arising from co-moving pulse (55) coupling with the $\nu
_{0,b} (\tau _0)$ function presented in Section 5; the dotted lines
denote all the lines in the upper panel of Fig. 11. Parameters are
the same as thy are in Fig. 12. The thin solid line stands for the
decaying phase of pulses and the thick solid line denotes the rising
phase.} \label{Fig. 18}
\end{figure}

For these two co-moving pulses, we explore the peak energy
evolutionary curve and the relation between the integrated flux and
energy. They are shown in Figs. 12 and 13 respectively. Fig. 12
shows that the skewness leads to a forward shifting of the peak of
EPC. It makes the offset of the peak of EPC larger when comparing
this peak with the time position of the peak count or peak flux (see
Fig. 4). The EPC in the decaying phase is obviously less affected by
the factors considered here. It is interesting that the shape of the
EPC is that of an incomplete pulse. The lack of the very small
values of $E_p$ in the rising phase must be due to the fact that the
co-moving pulses are assumed to start to emit at a high energy band.
This assumption seems to be reasonable, and therefore one could
expect the EPC of GRB pulses to possess an incomplete pulse shape if
these pulses are suffered from the curvature effect. Like what shown
in the EPC, the power law relation between the flux and peak energy
in the decaying phase is stubborn (see Fig. 13). The factors
concerned have no effects on this relation. On the contrary, the
relation in the rising phase varies significantly according to the
evolutionary pattern of intrinsic emission. As shown in Fig. 13,
although the curves in the rising phase are so different, they are
all on the right-hand-side of the decaying curve, which makes the
relation within the whole pulse interval a hook-like curve.

As proposed in Section 3, it will be natural if a co-moving pulse
experiencing a soft-to-hard spectral evolution phase (which might be
quite short) and then a hard-to-soft phase during a shock. One thus
can expect that the features of the curves studied in this section
might be common in the GRB pulses if they do arise from the emission
of a relativistically expanding fireball.

\section{Influence of the Lorentz factor}

In the above analysis, one important factor is not taken into
account, which is the Lorentz factor of the expanding motion of
fireballs. Would it give rise to entirely different results? We
explore this issue by adopting various values of the Lorentz factor
and then comparing the results. The peak energy evolutionary curve
and the relation between the integrated flux and peak energy are
investigated.

\begin{figure}[tbp]
\begin{center}
\includegraphics[width=5in,angle=0]{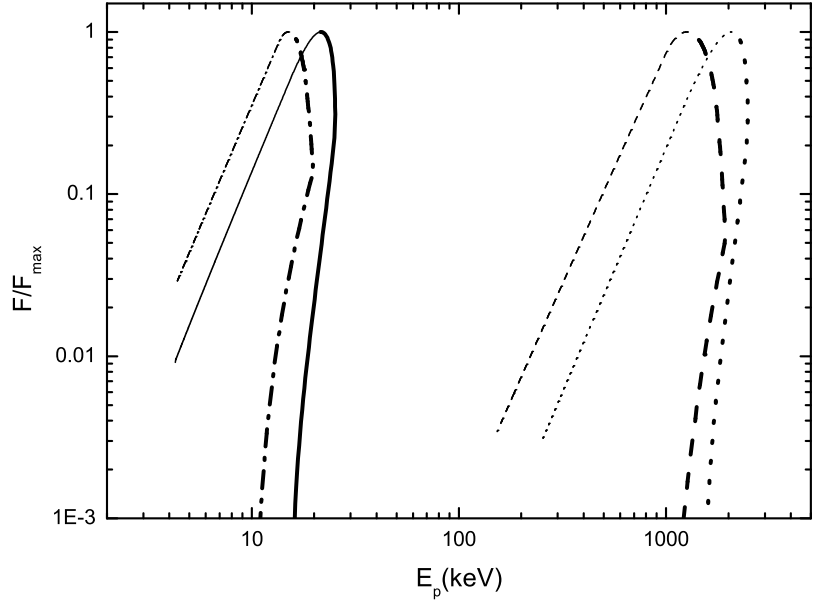}
\end{center}
\caption{Relation between the integrated flux and peak energy
arising from fireballs expanding with $\Gamma=10$ (the solid line
and the dashed dotted line) and $\Gamma=1000$ (the dashed line and
the dotted line) respectively. The dashed dotted line and the dashed
line stand for the curves associated with the revised broken power
law emission, co-moving pulse (55) coupling with the $\nu _{0,b}
(\tau _0)$ function presented in Section 5; the solid line and the
dotted line represent that associated with the modified Comptonized
radiation, co-moving pulse (56). Parameters other than the Lorentz
factor are the same as that adopted in Fig. 13. Thin lines stand for
the decaying phase of pulses and thick lines denote the rising
phase.} \label{Fig. 20}
\end{figure}

We adopt the intrinsic pulses discussed in last section to study
this issue, where, parameters other than the Lorentz factor are
maintained. The Lorentz factor $\Gamma=100$ is replaced with
$\Gamma=10$ and $\Gamma=1000$ respectively to show if this quantity
plays a role in the two relationships concerned. The analysis shows
that, although the adopted Lorentz factors are significantly
different, the features of EPC observed in Fig. 12 are maintained
(the figure is omitted). Shown in Fig. 14 are the relations between
the integrated flux and peak energy, calculated with the two Lorentz
factors. The shape of the curves of the integrated flux vs. peak
energy is the same as that shown in Fig. 13. The index of the power
law relation between the integrated flux and peak energy in the
decaying phase changes a little. In the case of the revised broken
power law emission, the index ranges from $\gamma =2.84$ (for
$\Gamma=10$) to $\gamma =2.99$ (for $\Gamma=1000$); and in the case
of the modified Comptonized radiation, the index changes from
$\gamma =2.95$ (for $\Gamma=10$) to $\gamma =3.07$ (for
$\Gamma=1000$). It seems that the larger the Lorentz factor, the
larger value the index. A noticeable feature revealed in Fig. 14 is
the shift of the peak energy range, which is expectable due to the
Doppler shifting.

\section{Signature of the curvature effect}

It is known that light curves arising from the emission of an
intrinsic $\delta$ function pulse with a mono-color radiation over
the whole fireball surface share the same profile, a marginal
decaying curve [17]. Light curves of real emission were found to
deviate from this curve by a reverse-S feature in their decaying
phase and this could serve as a signature of the curvature effect.
Here we consider a co-moving $\delta$ function pulse radiated at a
fixed energy (a mono-color radiation), trying to find if there
exists a similar curve in terms of the peak energy.

The co-moving $\delta $ function pulse with a mono-color radiation
is taken as
\begin{equation}
\widetilde{I}_{0,\nu }(\tau _0,\nu _0)=I_0\delta (\tau _0-\tau
_{0,0})\delta (\nu _0-\nu _{0,0})\qquad (0\leq \tau _{0,0}).
\end{equation}
Not losing the generality, we take $\tau _{0,0}=0$. When applying
equation (40) we take $\tau _{0,\min }=\tau _{0,\max }=0$ since
$\tau _0$ is confined by $\tau _{0,\min }\leq \tau _0\leq \tau
_{0,\max }$ and there is emission only at $\tau _0=\tau _{0,0}=0$.
We consider the emission from the whole fireball surface and then
take $\theta _{\min }=0$ and $\theta _{\max }=\pi /2$. Thus, from
(40) we get $ 0\leq \tau \leq 1$. The relation between the observed
frequency and the emitted rest frame frequency is formula (36).
Applying (57), the formula comes to
\begin{equation}
\nu (\tau )=\frac{\nu _{0,0}}{(1-\beta +\beta \tau )\Gamma }.
\end{equation}
This describes the curve of the development of $\nu $ that arises
from the emission of a co-moving $\delta $ function pulse radiated
with a mono-color spectrum.

One can check that the maximum of $\nu $ is reached when $\tau =0$.
That leads to $ \nu _{\max }=\frac{\nu _{0,0}}{(1-\beta )\Gamma }$.
Equation (58) then could be written as $ \frac{\nu (\tau )}{\nu
_{\max }}=\frac{1-\beta }{1-\beta +\beta \tau }$. Taking $\nu /\nu
_{\max }=1/2$ we get $ \tau _h=\frac{1-\beta }\beta $, which denotes
the time when the observed frequency is half of the maximum of $\nu
$. Combining these relations we get
\begin{equation}
\frac{\nu (\tau )}{\nu _{\max }}=\frac 1{1+ \tau/\tau _h}.
\end{equation}
It shows that, in terms of $\tau /\tau _h$, the curve $\nu /\nu
_{\max }$ is independent of the Lorentz factor.

Let us study various $E_p(t)$ curves discussed above in terms of
$t^{\prime }/t_h^{\prime }$, where $t^{\prime }$ is the observation
time set to the moment when the maximum of $E_p$, $E_{p,\max }$, is
observed and $ t_h^{\prime }$ denotes the time $t^{\prime }$ when
the observed $E_p$ is half of $ E_{p,\max }$. [Note that $t^{\prime
}/t_h^{\prime }=\tau/\tau _h$ since the coefficient in relating
$t^{\prime }$ and $\tau$ is canceled; see equation (35).] As shown
in Fig. 4, the spectral behavior in terms of $E_p$ in the decaying
phase of fireball pulses is quite stubborn. This is due to the fact
that the curvature effect dominates the evolution of the observed
spectrum in this period. We thus examine haw the evolutionary curve
of the peak energy in this period differs from the curve of (59)
which we call the marginal decaying form of the peak energy
evolutionary curve.

We compare the $E_p(t)$ curves in the decaying phase of the six
pulses discussed in Section 3 (see the caption of Fig. 2) with the
curve of (59) and find that the former well follow the latter, where
only very minute deviations are observed (the figure is omitted). It
is interesting that when a deviation is visible, the $E_p(t)$
evolutionary curve must possess a reverse-S feature relative to the
curve of (59). The result indicates that the marginal decaying form
(59) could serve as a signature of the curvature effect.

\begin{figure}[tbp]
\begin{center}
\includegraphics[width=5in,angle=0]{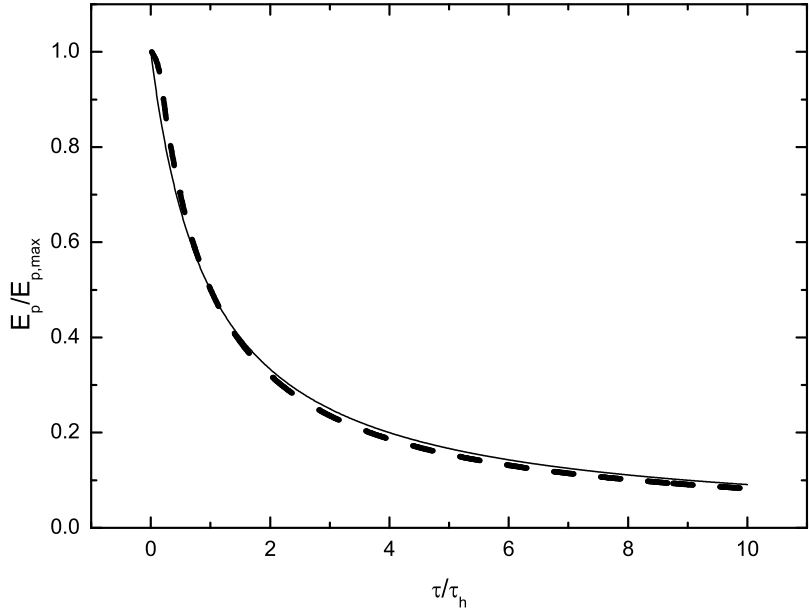}
\end{center}
\caption{Evolutionary curve (the dashed line) of the peak energy
$E_p$ in the decaying phase of a ``broad'' fireball pulse in the
case of the broken power law emission (55) with an unchanged
intrinsic spectrum (see the caption of Fig. 11). The solid line is
the curve of (59).} \label{Fig. 22}
\end{figure}

Do these conclusions hold in cases of other intrinsic emission forms
and other values of the Lorentz factor discussed above in Sections
4-6? The answer is yes. Presented in Fig. 15 is the peak energy
evolutionary curve arising from a ``broad'' co-moving pulse of the
broken power law emission (55) with an unchanged intrinsic spectrum
in the decaying phase of the pulse, which is seen to possess the
most deviation from the marginal decaying curve (59), detected from
the peak energy evolutionary curves of all the situations discussed
above. We find that the conclusions are not violated by all these
factors.

\section{Conclusions}

It is known that the power indexes in the relation between the flux
and peak energy derived from many GRBs are different from that
expected by the curvature effect, $\gamma \sim 3$. Some possible
interpretations for this variation were proposed by authors of [16]:
adiabatic and radiative cooling processes that extend the decay
timescale; a nonuniform jet; and the formation of pulses by external
shock processes. According to our analysis, the first one is
unlikely. The cooling time, long or short, is in fact included in
the decaying phase of co-moving pulses, and the decaying portion of
the pulses has little influence on the power index. Authors of [15]
proposed that the breaking of local spherical symmetry such as a
prolate or oblate shell geometry would result in a power law index
that differs from the spherical case. This interpretation is
somewhat similar to the second proposal of [16] and might be an
interesting outlet for the problem. In an investigation on the
hardness-intensity correlation (HIC) in GRB pulses, authors of [41]
found that some pulses exhibit a track jump in their HICs, in which
the correlation jumps between two power laws with the same index.
This was interpreted as a signature of the existence of strongly
overlapping pulses. This mechanism would naturally explain why the
index observed in some GRBs could be both larger and smaller than
what the curvature effect predicts. For example, an upper power law
line in Fig. 9 jumping to a lower power law line in the figure would
give rise to smaller index, while a lower line jumping to an upper
line would produce a larger index.

In the relation between the integrated flux and peak energy, the
following conclusions hold: in the decaying phase of pulses, the two
quantities are well related by a power law where the index is about
3, being free of the intrinsic emission and the Lorentz factor; the
relation in the rising phase differs significantly and the overall
relation between the two quantities varies and shifts enormously,
depending on the form or width of the intrinsic emission of pulses
and on parameters such as the Lorentz factor.

The pattern of the spectral evolution of the intrinsic emission of a
co-moving pulse within its rising phase could be observed in the
relation between the integrated flux and peak energy. For an
unchanged intrinsic spectrum, the relation in the rising phase is a
straight line in parallel with the axis of the flux; for a
hard-to-soft intrinsic spectrum, the flux decreases with the
increasing of the peak energy within this phase; for a
soft-to-hard-to-soft intrinsic spectrum, the flux generally
increases with the increasing of the peak energy in the rising
phase.

Besides the above conclusions, the following could also be reached
for pulses arising from a relativistically expanding fireball. a)
The spectrum of a pulse in its decaying phase differs slightly for
different intrinsic spectral evolution patterns, and hence it must
be dominated by the curvature effect. b) An intrinsic
soft-to-hard-to-soft spectral evolution within a co-moving pulse
would give rise to a pulse-like evolutionary curve for the peak
energy. c) There exists the marginal form of $1/(1+ \tau/\tau _h)$
for the peak energy evolutionary curve in the decaying phase of
pulses, and in many cases the peak energy evolutionary curve well
follows the form and when the former deviates from the latter it
deviates in a reverse-S way.

\acknowledgments

Supported by the National Natural Science Foundation of China (No.
10573005 and No. 10747001).

[1]{} Rees M J and Meszaros P 1994 {\it Astrophys. J.} {\bf 430} L93

[2]{} Burrows D N, Romano P, Falcone A, Kobayashi S, Zhang B,
Moretti A, O'Brien P T, Goad M R, Campana S, Page K L, Angelini L,
Barthelmy S, Beardmore A P, Capalbi M, Chincarini G, Cummings J,
Cusumano G, Fox D, Giommi P, Hill J E, Kennea J A, Krimm H, Mangano
V, Marshall F, M¨¦sz¨¢ros P, Morris D C, Nousek J A, Osborne J P,
Pagani C, Perri M, Tagliaferri G, Wells A A, Woosley S, Gehrels N
2005 {\it Science} {\bf 309} 1833

[3]{} Tagliaferri G, Goad M, Chincarini G, Moretti A, Campana S,
Burrows D N, Perri M, Barthelmy S D, Gehrels N, Krimm H, Sakamoto T,
Kumar P, M¨¦sz¨¢ros P I, Kobayashi S, Zhang B, Angelini L, Banat P,
Beardmore A P, Capalbi M, Covino S, Cusumano G, Giommi P, Godet O,
Hill J E, Kennea J A, Mangano V, Morris D C, Nousek J A, O'Brien P
T, Osborne J P, Pagani C, Page K L, Romano P, Stella L, Wells A 2005
{\it Nature} {\bf 436} 985

[4]{} Butler N R and Kocevski D 2007 {\it Astrophys. J.} {\bf 663}
407

[5]{} Nousek J A, Kouveliotou C, Grupe D, Page K L, Granot J,
Ramirez-Ruiz E, Patel S K, Burrows D N, Mangano V, Barthelmy S,
Beardmore A P, Campana S, Capalbi M, Chincarini G, Cusumano G,
Falcone A D, Gehrels N, Giommi P, Goad M R, Godet O, Hurkett C P,
Kennea J A, Moretti A, O'Brien P T, Osborne J P, Romano P,
Tagliaferri G, Wells A A 2006 {\it Astrophys. J.} {\bf 642} 389

[6]{} O'Brien P T, Willingale R, Osborne J, Goad M R, Page K L,
Vaughan S, Rol E, Beardmore A, Godet O, Hurkett C P, Wells A, Zhang
B, Kobayashi S, Burrows D N, Nousek J A, Kennea J A, Falcone A,
Grupe D, Gehrels N, Barthelmy S, Cannizzo J, Cummings J, Hill J E,
Krimm H, Chincarini G, Tagliaferri G, Campana S, Moretti A, Giommi
P, Perri M, Mangano V, LaParola V 2006 {\it Astrophys. J.} {\bf 647}
1213

[7]{} Zhang B, Fan Y Z, Dyks J, Kobayashi S, M¨¦sz¨¢ros P, Burrows D
N, Nousek, J A, Gehrels N 2006 {\it Astrophys. J.} {\bf 642} 354

[8]{} Zhang B-B, Liang E-W and Zhang B 2007 {\it Astrophys. J.} {\bf
666} 1002

[9]{} Qin Y-P, Su C-Y, Fan J H, Gupta A C 2006 {\it Phys. Rev. D}
{\bf 74} 063005

[10]{} Fenimore E E, Madras C D and Nayakshin S 1996 {\it Astrophys.
J.} {\bf 473} 998

[11]{} Sari P and Piran T 1997 {\it Astrophys. J.} {\bf 485} 270

[12]{} Kumar P and Panaitescu A 2000 {\it Astrophys. J.} {\bf 541}
L51

[13]{} Qin Y-P 2002 {\it Astro. Astrophys.} {\bf 396} 705

[14]{} Ryde F and Petrosian V 2002 {\it Astrophys. J.} {\bf 578} 290

[15]{} Kocevski D, Ryde F and Liang E 2003 {\it Astrophys. J.} {\bf
596} 389

[16]{} Dermer C D 2004 {\it Astrophys. J.} {\bf 614} 284

[17]{} Qin Y-P and Lu R-J 2005 {\it Mon. Not. Roy. Astron. Soc.}
{\bf 362} 1085

[18]{} Shen R-F, Song L-M and Li Z 2005 {\it Mon. Not. Roy. Astron.
Soc.} {\bf 362} 59

[19]{} Qin Y-P, Zhang Z-B, Zhang F-W, Cui X-H 2004 {\it Astrophys.
J.} {\bf 617} 439

[20]{} Qin Y-P, Dong Y-M, Lu R-J, Zhang B-B, Jia L-W 2005 {\it
Astrophys. J.} {\bf 632} 1008

[21] Zhang F-W, Qin Y-P 2005 {\it Chin. Phys.} {\bf 14} 2276

[22]{} Liang E W, Zhang B, O'Brien P T, Willingale R, Angelini L,
Burrows D N, Campana S, Chincarini G, Falcone A, Gehrels N, Goad M
R, Grupe D, Kobayashi S, M¨¦sz¨¢ros P, Nousek J A, Osborne J P, Page
K L, Tagliaferri G 2006 {\it Astrophys. J.} {\bf 646} 351

[23] Lu R-J, Qin Y-P and Zhang F-W  2007 {\it Chin. Phys.} {\bf 16}
1806

[24]{} Norris J P, Share G H, Messina D C, Dennis B R, Desai U D,
Cline T L, Matz S M, Chupp E L 1986 {\it Astrophys. J.} {\bf 301}
213

[25]{} Kargatis V E, Liang E P, BATSE Team 1995 {\it Astrophys.
Space Sci.} {\bf 231} 177

[26]{} Ryde F and Svensson R 2000 {\it Astrophys. J.} {\bf 529} L13

[27]{} Ryde F and Svensson R 2002 {\it Astrophys. J.} {\bf 566} 210

[28]{} Borgonovo L and Ryde F 2001 {\it Astrophys. J.} {\bf 548} 770

[29]{} Liang E W, Dai Z G and Wu X F 2004 {\it Astrophys. J.} {\bf
606} L29

[30]{} Yonetoku D, Murakami T, Nakamura T, Yamazaki R, Inoue A K,
Ioka K 2004 {\it Astrophys. J.} {\bf 609} 935

[31]{} Kargatis V E, Liang E P, Hurley K C, Barat C, Eveno E, Niel M
1994 {\it Astrophys. J.} {\bf 422} 260

[32]{} Band D 1997 {\it Astrophys. J.} {\bf 486} 928

[33]{} Band D, Matteson J, Ford L, Schaefer B, Palmer D, Teegarden
B, Cline T, Briggs M, Paciesas W, Pendleton G, Fishman G,
Kouveliotou C, Meegan C, Wilson R, Lestrade P 1993 {\it Astrophys.
J.} {\bf 413} 281

[34]{} Preece R D, Pendleton G N, Briggs M S, Mallozzi R S, Paciesas
W S, Band D L, Matteson J L, Meegan C A 1998 {\it Astrophys. J.}
{\bf 496} 849

[35]{} Preece R D, Briggs M S, Mallozzi R S, Pendleton G N, Paciesas
W S, Band D L 2000 {\it Astrophys. J. Suppl.} {\bf 126} 19

\end{document}